\documentclass{aa}  
\usepackage[]{natbib}

\usepackage{graphicx}
\usepackage[utf8]{inputenc}
\usepackage[breaklinks=true]{hyperref}
\usepackage{pdflscape}
\usepackage{afterpage}

\usepackage{mathtools}
\usepackage{txfonts}
\usepackage{xcolor}

\bibpunct{(}{)}{;}{a}{}{,} 

\newcommand{\Msun}{$M_{\rm{sun}}$}

\newcommand{\Ho}{$H_{0}$}

\newcommand{\Hunit}{$\rm{km\,s^{-1}\,Mpc^{-1}}$}
\newcommand{\hst}{{\it HST}}
\newcommand{\jwst}{{\it JWST}}

\newcommand{\kcorr}{$K-$correction}

\newcommand{\logg}{$\log{g}$}
\newcommand{\logP}{$\log{P}$}
\newcommand{\teff}{$T_{\mathrm{eff}}$}

\newcommand{\Aapp}{$A^{\mathrm{app}}$}

\newcommand{\Klam}{$K(\Lambda)$}
\newcommand{\Kw}{$K^{\mathrm{ext}}(W_{\mathrm{H,VI}})$}
\newcommand{\Ki}{$K^{\mathrm{ext}}(F814W)$}
\newcommand{\Kext}{$K^{\mathrm{ext}}(\Lambda)$}

\newcommand{\Kred}{$K^{\mathrm{ext}}_{\mathrm{red}}(\Lambda)$}
\newcommand{\Kredz}{$K^{\mathrm{ext}}_{\mathrm{red}}(\Lambda,z)$}
\newcommand{\Ktab}{$K^{\mathrm{ext}}$}

\begin{document}

   \title{Relativistic corrections for measuring Hubble's constant to $1\%$ using stellar standard candles}
    

   \author{Richard~I. Anderson}

  \institute{Institute of Physics, Laboratory of Astrophysics, \'Ecole Polytechnique F\'ed\'erale de Lausanne (EPFL), Observatoire de Sauverny, 1290 Versoix, Switzerland\\\email{richard.anderson@epfl.ch}}

   \date{Submitted: 26 June 2021; accepted: 8 November 2021}

  \abstract{We have estimated relativistic corrections for cosmic distance estimates based on stellar standard candles such as classical Cepheids and stars near the Tip of the Red Giant Branch (TRGB stars) with the goal of enabling a future unbiased $1\%$ measurement of Hubble's constant, \Ho. 
  We considered four effects: \kcorr s, time-dilation, the apparent change of host dust extinction due to non-comoving reference frames, and the change of observed color due to redshift.
  Using stellar model atmospheres, we computed extinction-dependent \kcorr s for a wide range of effective temperatures between $3500$ and $6000$\,K, iron abundances between $\mathrm{[Fe/H]}=-2.0$ and $0.5$, surface gravity between $\log{g}=2.0$ and $0.0$, and host reddening (up to $\mathrm{E(B-V)^{host}}=0.5$) for a range of redshifts corresponding to distances of $\sim 20-120$\,Mpc ($z$ between $0.005$ and $0.03$) in several \hst, \jwst, and 2MASS filters. 
  The optical-NIR Wesenheit function applied by the Cepheid distance ladder is particularly useful for limiting the magnitude of \kcorr s and for mitigating complications arising from host dust extinction. 
  Missing host extinction corrections related to the circumgalactic medium and circumstellar environments arising from stellar mass-loss are discussed as potential systematics of TRGB distance measurements. 
  However, their effect is estimated to be insufficient to explain differences in \Ho\ values based on Cepheids or TRGB supernova calibrations.
  All stellar standard candle observations require relativistic corrections in order to achieve an unbiased $1\%$ \Ho\ measurement in the future. Applying the \kcorr, the redshift-Leavitt bias correction, and a correction for the Wesenheit slope redshift dependence, the Cepheid-based \Ho\ measurement increases by $0.45\pm0.05$\,\Hunit\ to $H_0^{\mathrm{SH0ES}}=73.65 \pm 1.30$\,\Hunit, raising the tension with the early-Universe value reported by the {\it Planck} Collaboration from $4.2\sigma$ to $4.4\sigma$. 
  For TRGB-based \Ho\ measurements, we estimate a $\sim 0.5\%$ upward correction for the methodology employed by Freedman et al. ($H_0^{\mathrm{CCHP}} = 70.2\pm1.7$\Hunit) and an even smaller $-0.15\%$ downward correction for the methodology employed by Anand et al. ($H_0^{\mathrm{EDD}} = 71.4 \pm 1.8$\,\Hunit). The opposite sign of these corrections is related to different reddening systematics and reduces the difference between the studies by $\sim 0.46$\,\Hunit.
  The optical-NIR Wesenheit function is particularly attractive for accurate distance measurements because it advantageously combines measurements in filters where \kcorr s have opposite signs. The {\it JWST/NIRCAM} F277W filter is of particular interest for TRGB stars thanks to its insensitivity to (weak) host reddening and \kcorr s below the level of $1\%$ at Coma cluster distances.
  }

   \keywords{Distance scale -- Relativistic processes -- Methods: observational -- Stars: distances -- Stars: variables: Cepheids -- dust, extinction -- Galaxies: distances and redshifts }

   \maketitle
%

\section{Introduction}

Hubble's constant, \Ho, which quantifies the present-day expansion rate of the Universe, is intimately linked to the size and age of the Universe and therefore a fundamental quantity for cosmology. A high-accuracy prior of \Ho\ is required for constraining the equation of state of dark energy and to provide an absolute scale to many astrophysical and cosmological inquiries \citep[e.g.][]{Suyu2012,Weinberg2013}. More recently, \Ho\ has been hotly debated in terms of a \emph{tension} between values of \Ho\ derived from observations of the present-day Universe \citep[e.g.][]{Riess2016} and cosmology-dependent inferences based on early-Universe observations, such as the Cosmic Microwave Background \citep{Planck2020H0}, cf. \citet{Verde2019} and \citet{DiValentino2021} for reviews. As argued by \citet{Riess2020nature}, the two types of \Ho\ determinations provide a cosmological end-to-end test: any significant discrepancy between the observations of the early and present-day Universe would either indicate measurement error or incompleteness of the $\Lambda$CDM model used to interpret early-Universe phenomena, such as the cosmic microwave background (CMB) powerspectrum, the cosmological sound horizon, or Big Bang Nucleosynthesis. 

Type-Ia supernovae (SNeIa) play a particularly important role for measuring \Ho\ in the present-day Universe thanks to the great precision they afford in mapping relative distances along the Hubble flow at distances insensitive to cosmology. However, as relative distance indicators, SNeIa require absolute calibration provided by distance ladders composed of stellar standard candles (SSCs), such as classical Cepheid variable stars (henceforth: Cepheids) or stars near the Tip of the Red Giant Branch (henceforth: TRGB stars), and SNeIa. Such distance ladders allow a purely empirical measurement of \Ho\ tied to an absolute scale set by geometric or trigonometric measurements. The most precise distance ladder-based \Ho\ measurements now feature relative uncertainties of $\sigma_{H_0}/H_0 \approx 2\%$ or better \citep[e.g.][]{Riess2021,Freedman2021,Anand2021H0}. 

However, debate has ensued over the accuracy of the various SNeIa luminosity zero-point calibrations \citep[e.g.][]{Efstathiou2020}, sparking significant efforts to quantify and mitigate any intervening systematics related to parallax accuracy \citep{Riess2018,Riess2021,Breuval2020,Groenewegen2021}, photometric techniques employed 
\citep{Javanmardi2021,Anand2021H0}, reddening \citep[e.g.][]{Freedman2020,Gorski2020,Hoyt2021,Skowron2021,Soltis2021,Moertsell2021}, source blending \citep{AndersonRiess2018,Riess2020amp}, and chemical composition \citep{Gieren2018,Breuval2021}, among others. \citet{Anderson2019rlb} recently showed that redshift-Leavitt bias (RLB) due to time-dilation currently leads to a $0.3\%$ \Ho\ bias of \Ho\ measurements based on the Cepheid+SNeIa distance ladder. This bias grows beyond $1\%$ when seeking to measure more distant Cepheids and Mira stars, e.g., using the soon-to-be-launched \jwst. Since RLB leads to \emph{systematic underestimates} of \Ho, the true value of \Ho\ requires upward correction to higher values that tend to be more discrepant with {\it Planck}.

Among key systematics of SNeIa are the so-called \kcorr s initially proposed by \citet{Hubble1936K} that translate between the emitted and observed spectral energy distributions (SEDs).
The \kcorr\ in its modern form was first presented by \citet{Humason1956} who credited Merle F. Walker with having independently arrived at a similar result in the same form in 1948. Further historical information on the development of the \kcorr\ is found in \citet{Sandage1995}. \citet{Oke1968}, \citet{Hamuy1993}, and \citet{Kim1996} refined \kcorr s for application to SNeIa, and \citet{Hogg2002} presented an updated and didactic description, which supersedes previous discussions thanks to clear definitions and improved consistency. 
\citep{Nugent2002} considered the effect of \kcorr s and extinction effects on supernova light curves, albeit without incorporating reddening into the computation of $K$. 

\citet{McCall2004} investigated the redshift dependence of extinction corrections for redshifts far beyond the Hubble flow ($z = 0.4$). McCall further developed a methodology for correcting extinction incurred both in the emitter and the observer frame, pointing out the dependence of total-to-selective extinction on redshift.  \citet{McCall2004} also presented the first application of \kcorr s to Cepheid distances that sought to elucidate differences among distance measurements reported for NGC\,4258 and different regions of M31. McCall's main focus was the interplay between localized (emitter-frame) and foreground (Milky Way) extinction and how redshift changes the total-to-selective extinction required to correct distance measurements, however. Specifically, they neither discussed the impact of \kcorr s on the calibration of the SNeIa luminosity zero-point, nor on the measurement of \Ho.

Relativistic effects bias \Ho\ measurements because the expanding Universe increasingly separates the observer ($z = (\lambda - \lambda_0)/\lambda_0 \approx 0$) and emitter inertial frames (SSCs in SN-host galaxies, $z>0$) as a function of distance. Until a few years ago, the \Ho\ measurements were not sufficiently precise for this effect to be relevant, so that the assumption of comoving SSC calibration and application inertial frames was acceptable. However, the aim achieving a $1\%$ \Ho\ measurement to understand the origins and implications of the Hubble tension requires reassessment of systematic effects previously considered insignificant, including the assumption of comoving reference frames. Since redshift increases linearly with distance (at low $z$), there should exist a threshold distance beyond which relativistic effects cannot be ignored in the pursuit of a $1\%$ measurement. This is particularly true for the distance ladder for two principal reasons. Firstly, the \emph{calibration} of SSCs is performed in the very nearby Universe, including notably the Milky Way \citep[MW]{Cerny2020,Riess2021}, the Large and Small Magellanic Clouds \citep[LMC and SMC]{Riess2019,Hoyt2021}, and/or the megamaser galaxy NGC\,4258 \citep{Reid2019,Jang2021,Anand2021H0}, whereas SSCs are \emph{applied} to determine the distance to SN-host galaxies whose redshift relative to the calibration inertial frame is significantly non-zero \citep[e.g.][]{Riess2016,Freedman2019}. Secondly, the greatest gains for \Ho\ precision are anticipated by increasing the number of SN-host galaxies \citep[e.g.][]{Riess2021}, which implies observing SSCs at greater distances because the SNeIa rate is volume-limited. Hence, the most promising efforts underway to improve \Ho\ precision increase the sensitivity of \Ho\ measurements to relativistic effects. To understand and mitigate any related biases that could occur in this endeavor, we here investigate the impact of relativistic effects, specifically \kcorr s and RLB, as well as host dust extinction on SSCs in SN-host galaxies. In so doing, we aim to formulate simple mitigation strategies both for  previously reported measurements and the next-generation distance ladder to be constructed using \jwst. 

This \emph{article} is structured as follows. Section\,\ref{meth} presents the extinction-corrected \kcorr\ applicable to SSCs used for calibrating SNeIa (Sec.\,\ref{meth:Kcorr}), reminds the basics of the Wesenheit function, and presents the dependence of the Wesenheit function's slope, $R^W$, on redshift (Sec.\,\ref{sec:Wesenheit}). RLB corrections are recalled in Sec.\,\ref{sec:rlb} and Sec.\,\ref{sec:H0} subsequently describes the combined relativistic distance correction and  its application to \Ho. Section\,\ref{sec:data} presents the information used to estimate \kcorr s for the two SSCs under study: Cepheids and TRGB stars. Section\,\ref{sec:results} presents the results  for Cepheids (Sec.\,\ref{res:Cepheids}) and TRGB stars (Secs.\,\ref{res:TRGB}) based on synthetic spectrophotometry computed using stellar atmosphere models, a comparison of corrections applicable to the Cepheid methodology used by Riess et al. and the TRGB methodology employed by Freedman et al.  (Sec.\,\ref{sec:reddening}), an estimate of corrections applicable to the color-dependent TRGB methodology employed by the extragalactic distance database \citet{Anand2021edd}, and an assessment of the implications for \Ho\ (Sec.\,\ref{res:H0}). Section\,\ref{sec:disc} provides additional discussion concerning the choice of input information (Sec.\,\ref{disc:empiricalSED}), guidance of the trends to be expected for Mira and J-AGB stars (Sec.\,\ref{disc:Mira}), and dust extinction (Sec.\,\ref{disc:dust}). The final Sec.\,\ref{conclusions} summarizes the results. Tabulated results are provided in electronic form via the online appendix \ref{app:tables}.

\section{Methodology\label{meth}}

\subsection{The $K-$correction\label{meth:Kcorr}}
The $K-$correction accounts for the difference between the (observed) apparent magnitude of a source and the apparent magnitude of the same source in its comoving (emitter) inertial frame. The difference is caused by two effects: a systematic shift of the spectral energy distribution (SED) incident on the photometric filter (bluer parts of the emitted SED pass through the photometric filter), and  a dimming effect due to the fixed-width filter appearing compressed when viewed from the source. 
As explained in \citet{Anderson2019rlb}, we compute \kcorr s using the observed redshift, $z = z_{\mathrm{obs}} = (\lambda - \lambda_0)/\lambda_0$, of SN-host galaxies \citep[e.g. from][]{Huchra1992}, noting that $z$ combines various redshift contributions \citep{Calcino2017}.  
Following \citet{Hogg2002}, the $K-$correction is defined as
\begin{equation}
    m_R = M_Q + \mu_0 + K_{QR} 
    \label{eq:DM}
\end{equation}
where $R$ and $Q$ denote (possibly different) photometric filters in the rest-frames of the observer and emitter, respectively, $m$ apparent magnitude, $M$ absolute magnitude, and $\mu_0 = 5\log{(D_L/10\,\mathrm{pc})}$ the true distance modulus and its relation to luminosity distance $D_L$. For spectral flux densities, $f_\lambda$, expressed per unit wavelength \citep[Eq.\,12]{Hogg2002}:
\begin{multline} \label{eq:K}
K = - 2.5\log{\left[ \frac{1}{1+z} \right] } \\
 - 2.5\log{ \left[ \dfrac{\int{ \mathop{d\lambda_o} \lambda_o f_\lambda(\lambda_o) R(\lambda_o) }}{ \int{ \mathop{d\lambda_o} \lambda_o g_\lambda^R R(\lambda_o) } } \frac{ \int{ \mathop{d\lambda_e} \lambda_e g_\lambda^Q Q(\lambda_e) } }{ \int{ \mathop{d\lambda_e} \lambda_e f_\lambda([1+z]\lambda_e) Q(\lambda_e)  }}  \right] } \ ,
\end{multline}
with $g_\lambda$ the standard source spectral flux density per unit wavelength, $\lambda_o = (1+z)\lambda_e$ the relationship between wavelength in the observer (subscript $_o$) and emitter ($_e$) inertial frames, $R$ and $Q$ the photometric transmission curves (filters), and $\log$ signifying the 10-base logarithm throughout this work. All magnitudes are in the Vega system, computed using the Vega spectrum provided by the \texttt{pysynphot} library\footnote{file \texttt{alpha\_lyr\_stis\_008.fits}, available at \url{https://ssb.stsci.edu/cdbs/calspec/}} to represent $g_\lambda$. 

Equation\,\ref{eq:K} requires detailed knowledge of the \emph{intrinsic} source spectrum, $f_\lambda^0$, since $K_{QR}$ is tied to luminosity or bolometric magnitude \citep[cf.][Appendix B and \citealt{Hogg2002}]{Humason1956}. However, $f_\lambda^0$ is not usually known for stellar standard candles. Instead, the absolute magnitudes of SSCs are determined by \emph{calibration}, that is, by observing apparent magnitudes, converting them to an absolute magnitude scale using geometric distance measurements, $d$, and, where possible, correcting for dust extinction 
$A^{\mathrm{cal}}_\lambda$. With the exception of NGC\,4258, SSC calibration is typically limited to objects closer than 100\,kpc, that is, at distances where peculiar motion dominates over redshift due to cosmic expansion. We therefore make the simplifying assumption that SSCs are calibrated in the observer's frame. 

The \kcorr\ as defined in Eq.\,\ref{eq:K} does not account for dust extinction. Differently put, Eq.\,\ref{eq:K} holds if extinction can be perfectly corrected. Since this is not usually possible, and since different strategies for reducing sensitivity to extinction are employed for different SSCs, we considered three types of extinction in this analysis: $A^{\mathrm{cal}}$ the dust extinction applicable to SSC calibration, $A^{\mathrm{fg}}$ the extinction due to foreground dust, and $A^{\mathrm{host}}$ the extinction due to localized host in the emitter frame.
Extending Eq.\,\ref{eq:K} by $A^{\mathrm{cal}}$, $A^{\mathrm{fg}}$, and $A^{\mathrm{host}}$ yields:
\begin{equation}\label{eq:DMext}
\begin{split}
m_R = & M_Q + \mu_0 + A^{\mathrm{fg}} + A^{\mathrm{host}} + K_{QR} \\
 = & M_Q^{\mathrm{cal}} - A^{\mathrm{cal}} + \mu_0 + A^{\mathrm{fg}} + A^{\mathrm{host}} + K_{QR} \ .
\end{split}
\end{equation}
We apply dust extinction by multiplying continuous stellar flux densities with a particular reddening law and color excess value specified in E(B-V). Hence, the extinction operation includes both the change in source color (reddening) and the reduction in source intensity (dimming). 
\begin{equation}\label{eq:extinction}
\begin{split}
a^{\mathrm{cal}}_o  (\lambda_o) & \equiv f^{\mathrm{cal}}_o(\lambda_o)  / f^{\mathrm{true}}_o (\lambda_o)  \ , \\
a^{\mathrm{fg}}_o   (\lambda_o) & \equiv f^{\mathrm{fg}}_o(\lambda_o)  / f^{\mathrm{true}}_o (\lambda_o)  \ , \\
a^{\mathrm{host}}_e (\lambda_e) & \equiv f^{\mathrm{host}}_e(\lambda_e) / f^{\mathrm{true}}_e (\lambda_e)  \ , 
\end{split}
\end{equation}
where each value of $a(\lambda) = 10^{-0.4 A_\lambda}$ can be calculated for any filter assuming a reddening law when the color excess is known. The color excess in two filters $\Lambda_1$ and $\Lambda_2$ is $E(\Lambda_1 - \Lambda_2) = m^0_{\Lambda_1} - m^0_{\Lambda_2} - ( m_{\Lambda_1} - m_{\Lambda_2} )$, that is, the difference between intrinsic and apparent source color. Thus:
\begin{equation}\label{eq:r*ebv}
A_{\Lambda} = R_{\Lambda} \cdot E(\Lambda_1 - \Lambda_2) = \frac{A_{\Lambda}}{A_{\Lambda_1}-A_{\Lambda_2}} \cdot E(\Lambda_1 - \Lambda_2) \ ,
\end{equation}
where $\Lambda$ denotes the filter used to estimate the dimming and $\Lambda_{1,2}$ the filters used to determine color excess. The ratio $A_{\Lambda}/(A_{\Lambda_1}-A_{\Lambda_2})$ is specified by a reddening law \citep[e.g.][]{Cardelli1989,Fitzpatrick1999}, which itself depends on source color, particularly so in the case of broadband photometry. We opted to not compute $R_\Lambda$ as a function of intrinsic source color in order to most closely follow observational practice. Instead, we computed $R_\Lambda$ using a slightly reddened $E(B-V)=0.1$ spectrum of Vega adopting the \citet{Cardelli1989} reddening law for typical MW dust, i.e., $R_V = 3.1$.

Dust extinction affects \kcorr s in two instances: 1) when transforming the calibrated SED to the emitter frame and 2) when computing the observed (dust extincted) source in the observer frame. To achieve 1), single-filter calibrations of SSCs require extinction corrections to avoid distance bias. Thus, the true (extinction-corrected) SED in the emitter frame $f_\lambda^{\mathrm{true}}(\lambda_e)$ is assumed to be identical to:
\begin{equation}\label{eq:ftrue}
f_\lambda^{\mathrm{true}}(\lambda_e) = f_\lambda^{\mathrm{cal}}(\lambda_o) / a^{\mathrm{cal}}_o \ .
\end{equation}
Any systematic errors associated with this extinction correction will lead to a chromatic bias when transposing the calibrator source to the emitter frame. 

\begin{figure}
\centering
\includegraphics{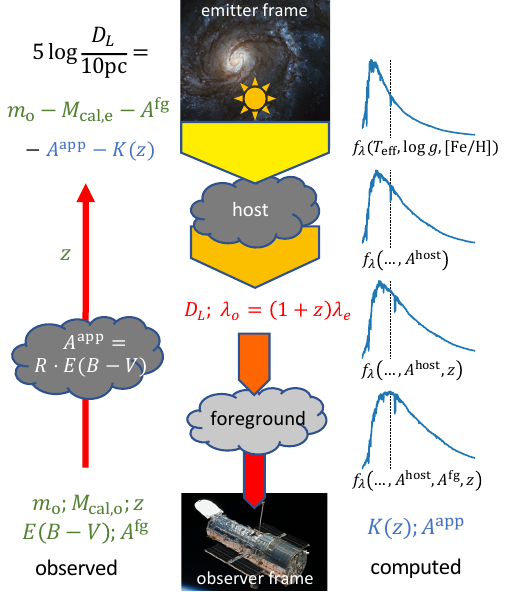}
\caption{\label{fig:comic}Cartoon illustration of the applied methodology. Center: the source is progressively reddened and attenuated by dust, redshift, and distance. Right: We successively apply host extinction, redshift, and foreground extinction to ATLAS9 SEDs before performing synthetic photometry in the observer frame to compute the `observed' SED used to compute the \kcorr\ and the apparent extinction correction based on apparent color excess. The dotted vertical lines indicate the position of the wavelength of the Ca\,II IR triplet in the emitter frame. Left: distance $D_L$ is determined using observed apparent magnitude $m_o$, observer-frame calibrated absolute magnitude $M_{\mathrm{cal}}$, dust extinction corrections based on apparent color excess (E(B-V)) and estimates of foreground dust extinction ($A_{\mathrm{fg}}$). Image credit for M100 and {\it HST}: NASA/STScI.}
\end{figure}

To achieve 2), the observed target flux density $f^{\mathrm{obs}}_\lambda(\lambda_o)$ is shaped by two attenuation events occurring in different inertial frames:
\begin{equation}\label{eq:fobs}
\begin{split}
f^{\mathrm{obs}}_\lambda(\lambda_o) = & \left[ f_\lambda^{\mathrm{true}}(\lambda_e) \cdot a^{\mathrm{host}}_e(\lambda_e) \right] \cdot a^{\mathrm{fg}}_o(\lambda_o) \\
= & f_\lambda^{\mathrm{true}}\left(\left[\frac{\lambda_o}{1+z}\right]\right) \cdot a^{\mathrm{host}}_o\left(\left[\frac{\lambda_o}{1+z}\right]\right) \cdot a^{\mathrm{fg}}_o(\lambda_o) \ .
\end{split}
\end{equation}
Hence, we computed $f^{\mathrm{obs}}_\lambda(\lambda_o)$ by successively applying the localized extinction in the emitter frame, blueshifting the SED to the observer frame, and applying foreground extinction. The center and right columns of Figure\,\ref{fig:comic} illustrate this.

For \kcorr\ computations applicable to single filter observations, we assumed that calibration observations of SSCs are accurately extinction-corrected, i.e., $a^{\mathrm{cal}}=1$, and that foreground correction is accurately determined by foreground dust maps, i.e., $a^{\mathrm{fg}}=1$. For all single-filter results presented, we thus exclusively vary $a^{\mathrm{host}}<1$ to test the impact of different amounts of localized dust extinction, expressed in the conventional units of $E(B-V)$ assuming the \citet{Cardelli1989} reddening law as explained above. Conversely for computations using the Wesenheit function, we adopted realistic sample averages for $\mathrm{E(B-V)^{host}}$ and $\mathrm{E(B-V)^{fg}}$ when computing $f_\lambda^{\mathrm{obs}}$ to remain sensitive to the interplay between redshift and the reddening law (cf. Sec.\,\ref{sec:Wesenheit}).

Starting from Eq.\,\ref{eq:DMext} and inserting Eqs.\,\ref{eq:extinction}, \ref{eq:ftrue}, and \ref{eq:fobs}, the \emph{extinction-dependent} \kcorr\ becomes:
\begin{equation}\label{eq:Kext}
\begin{split}
\MoveEqLeft
K^{\mathrm{ext}} = - 2.5\log{\left[ \frac{1}{1+z} \right] } \\
 - 2.5&\log{ \left[ \dfrac{\int{ \mathop{d\lambda_o} \lambda_o f^{\mathrm{obs}}_\lambda(\lambda_o) R(\lambda_o) }}{ \int{ \mathop{d\lambda_o} \lambda_o g_\lambda^R R(\lambda_o) } } \frac{ \int{ \mathop{d\lambda_e} \lambda_e g_\lambda^Q Q(\lambda_e) } }{ \int{ \mathop{d\lambda_e} \lambda_e f^{\mathrm{true}}_\lambda([1+z]\lambda_e) Q(\lambda_e)  }}  \right] } \ .
 \end{split}
\end{equation}

\Kext\ in Eq.\,\ref{eq:Kext} by definition depends on extinction.  While foreground and calibration extinction corrections are typically applied, direct measurements of $A^{\mathrm{host}}_{\Lambda}$ are difficult to obtain in practice due to observational limitations and because color excess as well as the ratio of total to selective extinction are affected by redshift \citep{McCall2004}. Yet, observations may yield an approximate estimate using an \emph{apparent color excess} calculated as the difference between the observed and the fiducial color of SSCs. Using this apparent color excess and assuming a reddening law, one may then compute an apparent extinction correction, $A^{\mathrm{app}}_\Lambda = R_\Lambda \cdot E^{\mathrm{app}}(B-V)$ to obtain an approximate correction for host extinction. 

Approximating $A^{\mathrm{host}} \approx A^{\mathrm{app}}$ and after applying foreground and calibration extinction corrections, Eq.\,\ref{eq:DMext} simplifies to:
\begin{equation}
m_R = M_Q^{\mathrm{cal}} + \mu_0 + K_{\mathrm{red},QR}^{ext}
\label{eq:Kred}
\end{equation}
with the \emph{extinction-dependent reduced $K-$correction}
\begin{equation}
K^{\mathrm{ext}}_{\mathrm{red}}(\Lambda) = K^{\mathrm{ext}}(\Lambda) - A^{\mathrm{app}}_\Lambda \ .
\label{eq:redKext}
\end{equation}
Of course, \Kred\ requires knowledge of the fiducial color. If there is no host extinction altogether, \Kred\ $=$ \Kext\ $= K$ after correcting for calibration and foreground extinction. 

The following text primarily discusses \kcorr s for Cepheids and TRGB stars applicable to the most recent methodologies and filter combinations, although the results computed span a much wider range of stellar parameters. For Cepheids, we mostly consider measurements using the near-infrared Wesenheit function, which is reddening-free by construction (cf. Sec.\,\ref{sec:Wesenheit}). Thus, we adopt realistic values for calibration, foreground, and host extinction to calculate $K^{\mathrm{ext}}(\mathrm{H,VI}) = K^{\mathrm{ext}}(H) - R^W_{\rm{H,VI}}\cdot\left( K^{\mathrm{ext}}(V) - K^{\mathrm{ext}}(I) \right)$. Cepheid observations using individual passbands are discussed in terms of \Kred, that is, host extinction is corrected using the apparent color excess, cf. the tables in the online appendix that contain $K^{\mathrm{ext}}(\Lambda)$ and $A^{\mathrm{app}}_\Lambda$ separately for a broad range of $\mathrm{E(B-V)^{host}}$ values.

For TRGB stars, we focus on $K^{\mathrm{ext}}(\mathrm{F814W})$ assuming that calibration and foreground extinction is corrected, so that $\mathrm{E(B-V)^{fg} = E(B-V)^{cal}} = 0$, and do not apply extinction corrections based on apparent color excess.  The effect of neglecting small amounts of host extinction among TRGB stars are discussed in Secs.\,\ref{sec:reddening} and \ref{disc:dust}.

\subsection{The Wesenheit function\label{sec:Wesenheit}}

The so-called {\it Wesenheit} function \citep{Bergh1975,Madore1976} is commonly employed to reduce extinction-related issues of distance measurements thanks to its \emph{reddening-free construction} for a given reddening law. Following \citet{Madore1982} the Wesenheit function is defined as: 
\begin{equation}\label{eq:Wesenheit}
m^W_{\Lambda_1,\Lambda_2\Lambda_3} = m(\Lambda_1) - R^W_{\Lambda_1,\Lambda_2\Lambda_3} \cdot \left( m(\Lambda_2) - m(\Lambda_3) \right) \ , \mathrm{ where}
\end{equation}
\begin{equation}\label{eq:RWesen}
R^W_{\Lambda_1,\Lambda_2\Lambda_3} = \dfrac{A_{\Lambda_1}}{A_{\Lambda_2} - A_{\Lambda_3}} 
\end{equation}
is the total-to-selective extinction ratio as given by a reddening law, normalized to a specific value of $R_V$, for any combination of filters $\Lambda_1$, $\Lambda_2$, $\Lambda_3$. Specifically, the near infrared (NIR) Wesenheit function used to calibrate the Cepheid distance ladder \citep{Riess2016,Riess2021} uses $m^W_{\mathrm{H,VI}} = m_H - 0.386\cdot(V-I)$, where $H$, $V$, and $I$ refer to Vega magnitudes in {\it HST/WFC3} F160W, F555W, and F814W filters, respectively. According to \citep{Riess2016}, the value of $R^W_{\mathrm{H,VI}} = 0.386$ corresponds to a \citet{Fitzpatrick1999} reddening law at $R_V = 3.3$. Different reddening laws yield different values of $R^W_{H,VI}$ for the same combination of filters and $R_V$, so that $R^W_{\mathrm{H,VI}}$ depends on the assumed reddening law. Moreover, $R^W_{\mathrm{H,VI}}$ depends on the SED of the source. For example, we find $R^W_{\mathrm{H,VI}} \sim 0.386$ for  slightly reddened Vega spectrum assuming the \citet{Fitzpatrick1999} reddening law and $R_V = 3.3$. For a $\sim 10$\,d MW Cepheid at the center of the instability strip (cf. Sec.\,\ref{data:Cepheids}), reddened by $E(B-V)=0.4$, and assuming $R_V=3.3$, the \citet[F99]{Fitzpatrick1999} reddening law yields $R^W_{H,VI} \sim 0.409$. Adopting instead the $R_V=3.1$ \citet[CCM]{Cardelli1989} reddening law yields $R^W_{\mathrm{H,VI}} = 0.464$. Since dust extinction is a chromatic process sensitive to redshift \citep{McCall2004}, $R^W$ strictly speaking only applies in the observer's frame and must in principle be redshifted to the emitter frame when measuring distance to avoid a redshift-related reddening-law bias. Computing $R^W_{\mathrm{H,VI}}$ for the fiducial 10\,d Cepheid yields:
\begin{equation}\label{eq:Rofz}
\begin{split}
\MoveEqLeft \mathrm{CCM}: & R^W_{\mathrm{H,VI}}(z) & =  0.464 + 0.126\, z \\
\MoveEqLeft \mathrm{F99}: & R^W_{\mathrm{H,VI}}(z) & =  0.408 + 0.105\, z \\
\end{split}
\end{equation}
The linear formulation of Eq.\,\ref{eq:Rofz} is adequate at low redshift (e.g. $z <0.03$); the relation becomes increasingly non-linear at larger $z$. 

Since the Wesenheit slope $R^W$ depends on redshift, and with $\Delta R^W(z) = R^W(z>0) - R^W(0)$, one may define a corrective term to apply to the absolute Wesenheit magnitude in the emitter frame:
\begin{equation}
\begin{split}
\label{eq:deltaMW}
\Delta M^W(z) & = M^{W}(z > 0) - M^{W}(z = 0) \\
  & = \left( R^W(z = 0) - R^W(z > 0) \right)\cdot (\Lambda_2 - \Lambda_3) \\
  & = - \Delta R^W(z) \cdot (\Lambda_2 - \Lambda_3) \ .
\end{split}
\end{equation}
Thus, $\Delta M_{\mathrm{H,VI}}^W(z) \approx - 0.126 \cdot z \cdot (V-I)$ for the CCM reddening law, and $-0.105 \cdot z \cdot (V-I)$ for F99. Since $\Delta M^W(z) < 0$ ($\Delta R^W(z) > 0$, $z>0$, and $(V-I)>0$), the absolute Wesenheit magnitudes of Cepheids in SN-host galaxies are intrinsically slightly brighter than previously thought. Expressed as a bias correction for distances based on the Wesenheit function, we obtain:
\begin{equation}
\begin{split}
\Delta \mu^{R^W}(z) & = \mu - \mu_0 \\
 & = m^W - M^{W}(z > 0) - \left( m^W - M^{W}(z=0) \right) \\
 & = - \Delta M^W(z) \\
 & = \Delta R^W(z) \cdot (\Lambda_2 - \Lambda_3) 
\end{split}
\label{eq:deltaMuRw}
\end{equation}
so that $\Delta \mu^{R_{\mathrm{H,VI}}^W}(z) \approx 0.126 \cdot z \cdot (V-I)$ for the CCM reddening law.

At $z=0.0245$, $\Delta \mu^{R^W_{\mathrm{H,VI}}}(z) \approx 3$\,mmag for $(V-I) \approx 1.0$\,mag, causing a $0.15\%$ systematic distance error at 100\,Mpc. The period-color relation of Cepheids further implies a small change of mean \teff\ as a function of pulsation period $P$, resulting in a minute period-dependence of observed Leavitt law (LL, Period-luminosity relation) slopes in galaxies, depending on the range of Cepheid periods observed.

\subsection{Redshift-Leavitt bias due to time dilation\label{sec:rlb}} 
\citet{Anderson2019rlb} described how time-dilation between non-comoving inertial frames biases cosmological distance measurements based on Leavitt laws of variable stars. In essence, the observed period $P_o$ is a clock observed to tick slower than its true period in the emitter frame, $P_e$, and thus: $P_o > P_e$. The difference in $\log{P}$ expected between the anchor galaxies belonging to the observer's rest frame, and the receding SN-host galaxies  is therefore $\Delta \log{P} = \log{P_o} - \log{P_e} = \log{[1+z]}$. For stars following a linear LL of the form $M = a + b\log{P}$, the resulting redshift-Leavitt bias (RLB) leads to overestimated luminosity and an absolute magnitude bias $\Delta M^{\mathrm{RLB}}(z) = M_o - M_e = b\log{[1+z]}$, which in distance modulus is expressed as:
\begin{equation}
    \Delta \mu^{\mathrm{RLB}}(z) = - b \log{[1+z]} \ .
\label{eq:RLB}
\end{equation}
Since longer period stars are brighter ($b < 0$) and redshift increases with distance ($z > 0$), the sign of the bias correction is generally positive ($\Delta \mu_{\mathrm{RLB}} > 0$). Hence the true distances ($\mu_0$) are shorter than apparent distances measured without RLB-correction, and RLB-corrections tend to \emph{increase} the value of $H_0 = v/D_L$.

\subsection{Effect of relativistic corrections on the Hubble constant\label{sec:H0}}

Inserting Eqs.\,\ref{eq:RLB} and \ref{eq:deltaMuRw} in Eq.\,\ref{eq:DMext} yields the combined relativistic distance correction:
\begin{equation}
\label{eq:deltaMuRel}
\begin{split}
\Delta \mu^{\mathrm{rel}}(z) & = \mu^{\mathrm{obs}} - \mu_0 \\
 & = K^{\mathrm{ext}} + \Delta \mu^{\mathrm{RLB}}(z) + \Delta \mu^{R^W} \\
 & = K^{\mathrm{ext}} - b \log{[1+z]} + \Delta R^W(z) \cdot (\Lambda_2 - \Lambda_3) \ , 
\end{split}
\end{equation}
where $\mu^{\mathrm{obs}}$ is the distance modulus measured without applying relativistic corrections, and $\mu_0$ the true distance modulus. Although both the RLB and Wesenheit slope corrections are small unto themselves, we note that they both increase $\Delta \mu^{\mathrm{rel}}$ since the LL slope $b < 0$ and $\Delta R^W(z) > 0$. Depending on the sign of \Kext, $\Delta \mu^{\mathrm{rel}}(z)$ is thus either brought closer to $0$ by the other two effects, or is further driven towards larger (positive) values. Of course, RLB does not apply to TRGB stars. Additionally, $\Delta R^W(z) = 0$ when the Wesenheit function is not used. Thus, for the Wesenheit function applied to Cepheids, $\Delta \mu^{\mathrm{rel}}(z)  = K^{\mathrm{ext}} - b \log{[1+z]} + \Delta R^W(z) \cdot (V-I)$. For TRGB stars observed in F814W, $\Delta \mu^{\mathrm{rel}}(z)  = K^{\mathrm{ext}}$; if reddening corrections based on apparent color excess are included  $\Delta \mu^{\mathrm{rel}}(z)  = K^{\mathrm{ext}}_{\mathrm{red}}$. For TRGB distances by \citet{Anand2021edd,Anand2021H0} employing the color calibration by \citet{Rizzi2007}, we additionally considered the effect of redshifted colors in Sec.\,\ref{sec:EDD}.

Using Eq.\,\ref{eq:deltaMuRel}, the correction to apply to literature luminosity distance estimates, $D_L$, is
\begin{equation}
    D^{\mathrm{true}}_L = D^{\mathrm{obs}}_L \cdot 10^{-0.2\Delta \mu^{\mathrm{rel}}(z) } \ .
    \label{eq:disttrue}
\end{equation}

For reported \Ho\ measurements based on distance ladders whose base calibration is provided by SSCs \citep[e.g.][]{Riess2016,Freedman2019}, the correction then becomes:
\begin{equation}
    H_0^{\mathrm{true}} = H_0^{\mathrm{obs}} \cdot 10^{0.2\Delta\mu^{\mathrm{rel}}(z)} \ .
    \label{eq:H0true}
\end{equation}
Depending on the sign of $\Delta\mu^{\mathrm{rel}}(z) > 0$, reported values of \Ho\ may require either upward or downward revisions, which would impact the significance of the discord between the present-day and early-Universe \Ho\ determinations.

\section{Fiducial information used for Cepheids and TRGB stars\label{sec:data} }

The formalism outlined in Sec.\,\ref{meth} requires detailed knowledge of stellar spectral energy distributions over a wide wavelength interval, in particular when considering Wesenheit magnitudes. To investigate the interplay of \kcorr s with stellar parameters, reddening, and redshift, we opted to base this investigation on the library of synthetic ATLAS9 stellar model atmosphere grid included in the python library \texttt{pysynphot}\footnote{\url{https://pysynphot.readthedocs.io/en/latest/}} \citep{pysynphot,Castelli2003}. The parameter range covered by these models (\teff$>3500$\,K, \logg$<0$) allowed us to investigate Cepheids and TRGB stars, and the following subsections describe the parameters adopted to represent these stars. Unfortunately, limitations of the ATLAS9 models precluded the computation of \kcorr s for cooler SSCs, such as oxygen-rich Mira variables \citep[e.g.][]{Whitelock2008,Huang2020} and the recently introduced $J-$region AGB stars \citep{Madore2020,Zgirski2021}. Some extrapolations of our results to these promising SSCs are discussed in Sec.\,\ref{disc:Mira}. Additional considerations concerning the use of models over empirical SEDs are presented in Sec.\,\ref{disc:models}.

All photometric transmission curves (filters) were downloaded from the Spanish Virtual Observatory Filter Profile Service\footnote{\url{http://svo2.cab.inta-csic.es/svo/theory/fps}}\footnote{\url{http://ivoa.net/documents/Notes/SVOFPS/index.html}}. We computed results for {\it HST/WFC3} filters F555W, F814W (UVIS channel), and F160W (IR channel), as well as {\it HST/ACS} filter F606W, since these are extensively used to measure \Ho\ to great precision \citep{Riess2018,Riess2021,Freedman2019}. Other filters are straightforward to include and we ensured that differences between F555W and F814W yield negligible differences among {\it HST/WFC3} and {\it HST/ACS}. In anticipation of this year's launch of the James Webb Space Telescope ({\it JWST}), we included the broadband {\it NIRCAM} filters F115W, F150W, F200W, and F277W. 2MASS filters $J$ and $K_s$ are also included for reference. In the context of the Wesenheit function, {\it HST} filters are occasionally abbreviated to $H$ (F160W), $V$ (F555W), and $I$ (F814W) band for better legibility.

\subsection{Classical Cepheids\label{data:Cepheids}}
\begin{figure}
\includegraphics{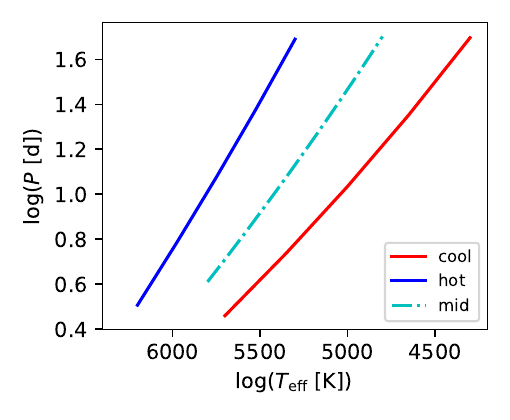} 
\caption{Dependence of \logP\ on \teff\ determined based on the pulsation analysis of Geneva stellar evolution models for Solar metallicity for Cepheids with typical initial rotation ($\omega=0.5$) \citep{Anderson2016rot}. The blue dash-dotted line represents the mid-point between the hot (BE) and cool (RE) edges of the instability strip, and we average over the second crossing. \label{fig:logPvsT}\label{fig:PCrelation}}
\end{figure}

Classical Cepheids are pulsating yellow (super-)giant stars whose spectral types range from mid-F to early-K. Their chromatic pulsations cause \teff\ to vary with pulsation phase at an amplitude frequently exceeding $1000$\,K \citep[e.g.][]{Kovtyukh2005}. Cepheids obey Leavitt laws \citep[LL, period-luminosity relation]{Leavitt1912} that allow the inference of luminosity $L$ from the straightforward period ($P$)measurement. Cepheid pulsations are driven primarily by the $\kappa$\,Mechanism, which requires a particular stellar structure to operate and confines Cepheids to a near-vertical strip in the Hertzsprung-Russell-Diagram (HRD). This narrow region, the so-called classical Instability Strip (IS), has a small finite width in \teff\ at fixed $L$. Since the IS is furthermore sloped towards cooler temperatures at higher $L$, there exists also a period-color relation (or an $L-$\teff relation) so that hotter Cepheids of period $P$ are more luminous than their cooler cousins with identical $P$ due to the Boltzmann law. Classical (linear) LLs marginalize over such effects, leading to intrinsic LL scatter due to several astrophysical processes  \citep[e.g.][]{Anderson2016rot}. The range of \teff\ at any given $P$ is up to several hundred Kelvins, cf. Fig.\,\ref{fig:logPvsT}. However, it is important to note that the IS is not homogeneously sampled due to the varying evolutionary timescale along the blue loop evolution, which makes it more likely to observe Cepheids closer to the hot IS boundary compared to the cool IS boundary. Given the uncertainties of the absolute position of the IS, we opted to consider Cepheids at the center of the IS.

We determined \teff\ at the center of the classical IS using the published pulsation analysis of Geneva stellar evolution models \citep[and \citealt{Georgy2013} and \citealt{Ekstroem2012} for model details]{Anderson2016rot}.  For Solar metallicity ($Z=0.014$) Cepheids, we found 
\begin{equation}
\log{(T_{\mathrm{eff}}\ \mathrm{[K]})} = -0.0753\cdot\log{P} + 3.8094 \ .
\label{eq:Teff_from_logP}
\end{equation} 
Thus, a $10$\,d Solar metallicity Cepheid at the center of the IS is expected to have \teff$=5421$\,K. This is close to the mean \teff$=5340$\,K of 22 MW Cepheids with $9.4 < P < 10.4$\,d whose temperatures have been determined by fitting photometrically sampled SEDs \citep{Groenewegen2020}.
At lower metallicities, the instability strip boundaries become hotter and steeper in the HRD. For lower metallicity models, we found $\log{T} = -0.0608\cdot\log{P} + 3.8043$ (LMC, $Z=0.006$) and $\log{T} = -0.0568\cdot\log{P} + 3.8073$ (SMC, $Z=0.002$).

Since Cepheids also obey a period-radius relation, we fitted a crude relation between cycle-averaged \logg\ and \logP\ based on measurements of 14 Galactic Cepheids presented by \citet{Proxauf2018} and found
\begin{equation}
\log{g} = -0.8808\cdot\log{P} + 2.1909 \ .
\label{eq:logg_from_logP}
\end{equation} 
For a 10\,d MW Cepheid, Eq.\,\ref{eq:logg_from_logP} yields \logg$\approx 1.3$. In line with these empirical values, we adopted for the fiducial 10\,d MW Cepheid model an ATLAS9 atmosphere with the parameters \teff$=5400$\,K, \logg$=1.5$, and [Fe/H]$=0.0$. 

For computations using the Wesenheit function, we adopted a mean color excess $E(B-V) = 0.4$ for computing $A^{\mathrm{cal}}$; this corresponds to the mean value \citep[$0.40 \pm 0.02$\,mag]{Groenewegen2018} of the 75 MW Cepheids for which {\it HST} photometry has been collected \citep{Riess2018,Riess2021}. We further adopted $E(B-V)^{\mathrm{fg}} = 0.0223$\,mag for computing $A^{\mathrm{fg}}$ in the case of Wesenheit magnitudes to match the mean foreground reddening \citep{Schlafly2011} of the 19 SN-host galaxies in \citet{Riess2016}. Different values ranging from $E(B-V)=0$ to $0.5$\,mag were used to investigate host reddening, covering a range of twice the spectroscopically measured reddening of a blue supergiant star in NGC\,4258 \citep[$E(B-V) = 0.23 \pm 0.03$\,mag]{Kudritzki2013}. Selecting SN-host galaxies with low inclinations, low reddening, and applying selection cuts to Cepheid candidates (e.g. mean color), is expected to constrain the range of true host reddenings to within the range adopted here.

\subsection{Stars near the TRGB\label{data:TRGB}}
The TRGB coincides with a well-defined evolutionary stage of low-mass stars before the He flash \citep[e.g.][]{Salaris1997trgb,Serenelli2017}. It has been used extensively to measure distances to hundreds of galaxies within $\sim 10$\,Mpc \citep[e.g.][]{Rizzi2007,Jacobs2009,Anand2021edd} and to calibrate the SNeIa luminosity zero-point for measuring \Ho\ \citep[e.g.][]{Lee1993,Jang2017,Freedman2019,Freedman2021,Anand2021H0}. 

The magnitude of the TRGB is determined by a marked decrease in the luminosity function along the red giant branch, located approximately at an absolute F814W  magnitude of $-4.06$\,mag \citep[e.g.][]{Cerny2020,Jang2021,Hoyt2021}. Near the $I-$band, and for a restricted range of parameters, the TRGB is roughly flat in color, whereas its slope changes from decreasing with color for shorter-wavelength observations to increasing with color for longer-wavelength observations \citep[e.g.][]{Valenti2004,Freedman2019,McQuinn2019,Madore2020}. The basis for using the TRGB as a standard candle is that the luminosity of RGB stars cannot exceed a certain luminosity without triggering very rapid ignition of He burning in the degenerate core, which displaces stars towards higher temperature and lower luminosity onto the red clump (higher metallicity) or the zero-age horizontal branch (ZAHB, lower metallicity populations). Details concerning the dependence of TRGB luminosity on chemical composition, core convection, age, etc. are found e.g. in \citet{Cassisi2017,Serenelli2017}. According to stellar evolution models, the TRGB's HRD position depends sensitively on metallicity, moving to hotter temperatures for lower metallicity, as well as age. Observationally, these issues can be mitigated either by focusing exclusively on low-metallicity stellar populations older than several ($\gtrsim 4$) Gyr \citep{Freedman2020}, or by correcting the absolute magnitude using a color term \citep{Rizzi2007}.

We consulted literature on spectroscopic studies of the RGB to determine the properties of a fiducial star at the TRGB. \citet{Lemasle2014} measured spectroscopy of metal-poor red giant stars in Fornax where the TRGB is clearly distinguishable. Their parameters for star `m0714' near the TRGB are \teff$=4090$\,K, \logg$=0.50$, [Fe/H]=$-1.80$; for the somewhat more metal-rich star `m0631' Lemasle et al. (expectedly) found the lower temperature \teff$=4035$\,K, \logg$=0.55$, [Fe/H]=$-0.85$. The study of rotational broadening along (presumably even more metal-rich) field red giants by \citet{Cortes2009} placed the TRGB near $\log{g} \approx 0.2$ and \teff$\approx 3890$\,K.

As a cross-check and to reduce the effect of high-metallicity interlopers, we also determined the parameters for the fiducial TRGB stars by consulting synthetic populations based on PARSEC isochrones computed via the CMD\footnote{\url{http://stev.oapd.inaf.it/cgi-bin/cmd_3.4}} webtool V3.4 \citep{Bressan2012,Chen2019}. We adopted the parameters of $\omega$\,Cen from \citet{Soltis2021} to compute 5 separate old and metal-poor ($11.5$\,Gyr, [Fe/H]$=-1.7$] populations of $10^5$\,\Msun. The most luminous stars in the $I$, $V-I$ color-magnitude diagrams based on the retrieved synthetic populations suggest \teff$=4300$\,K, $\log{g}=0.5$, and [Fe/H]$=-1.7$ as the fiducial parameters for TRGB stars. For the same age ($11.5$\,Gyr) and higher metallicity ($\mathrm{[M/H]}=-0.7$), these parameters shift to \teff$=3850$\,K, in line with the spectroscopic results mentioned above. We thus average between the theoretical TRGB star and the `m0714' star in \citet{Lemasle2014} and adopt \teff$=4200$\,K, $\log{g}=0.5$, and [Fe/H]$=-1.75$ for the fiducial stellar parameters at the TRGB.

Dust extinction is explicitly considered during the absolute magnitude calibration of the TRGB. Specifically, the recent \Ho\ measurement reported by \citet{Freedman2021} relied on TRGB absolute magnitude calibrations from four different sets of data, including MW Globular Clusters \citep[reddening corrected based on the literature compilation of color excesses in \citealt{Harris1996} (2010 edition)]{Cerny2020}, the Small and Large Magellanic Clouds \citep[reddening correction based on the \citealt{Skowron2021} reddening maps derived using red clump stars]{Hoyt2021}, and the megamaser host galaxy NGC\,4258 \citep[reddening corrected using all-sky maps]{Jang2021}.
The prior \Ho\ measurement by \citet{Freedman2020} had relied exclusively on a separate LMC absolute magnitude calibration where multi-band photometry was used to correct extinction effects directly without resorting to external measurements.  
Semi-theoretical reddening maps towards the Magellanic Clouds recently presented by \citet{Nataf2021} are generally consistent with the empirical maps by \citet{Skowron2021}, who also provided an in-depth discussion of pros and cons of various reddening map determinations.

We note a potentially interesting conceptual difference between the reddening corrections applied to the TRGB calibration in the Magellanic Clouds versus the calibration based on the TRGB in NGC\,4258 and the MW globular clusters. The reddening estimates in the Magellanic clouds are based on well-traceable CCD photometry of red clump stars, which are in a similar evolutionary state as the stars near the TRGB and exhibit similar colors. Conversely, the reddening estimates of MW globular clusters in \citet[2010 edition]{Harris1996} are not readily traceable in detail, and the online bibliography\footnote{\url{https://physics.mcmaster.ca/~harris/mwgc.dat}}\footnote{\url{https://physics.mcmaster.ca/~harris/mwgc.ref}} states that most reddening estimates are averages of color-magnitude diagram-based reddenings reported in the 1980s \citep{Webbink1985,Zinn1985,Reed1988}. The reddening correction to the TRGB in NGC\,4258 is based on the \citet{Schlafly2011} recalibration of the \citet{Schlegel1998} all-sky reddening maps and thus does not use stellar standard crayons at all. 
Despite these systematic differences among the calibrating sets, the following considers extinction to be accurately corrected during absolute calibration so that $\mathrm{E(B-V)^{cal}} = 0$.

Foreground dust extinction is usually considered explicitly based on all-sky reddening maps. We thus consider that foreground extinction is adequately removed and set $\mathrm{E(B-V)^{fg}} = 0$.

Localized extinction in SN-host galaxies (host extinction) is not usually considered and is indeed explicitly neglected when measuring distances via the TRGB method. As mentioned repeatedly by \citet{Freedman2019,Freedman2020,Freedman2021}, galactic halos are assumed to be \emph{entirely devoid} of gas and dust, and hence, of any extincting material because observations point to very low levels of extinction in galaxy halos. However, any amount of gas or dust leads to net extinction effect, which formally invalidates this assumption and introduces a (small) distance bias, since the systematic uncertainty due to host extinction is not Gaussian with zero mean. 

Host extinction of relevance to TRGB stars can arise from two main sources. Firstly, the circumgalactic medium (CGM) extends well into galaxy halos with contributions from inflowing gas as well as feedback from the supermassive black holes at the centers of galaxies \citep[for reviews on the subject, cf.][]{Putnam2012,Tumlinson2017}, and the composition and origin of the CGM is a very active subject of research for galaxy evolution \citep[e.g.][]{Anand2021cgm}. The typical color excess due to the CGM has been measured to be E(B-V)$\approx 0.01$\,mag within a few tens of kpc from the galaxy center along the semimajor axis \citep[e.g.][]{Zaritsky1994,Menard2010,Peek2015}. Overall, CGM reddening is most efficiently avoided by observing the TRGB in the outer halos, where the detection of the TRGB is also generally much cleaner \citep{Jang2021}.

Localized extinction can also arise from circumstellar environments fed by mass-loss during RGB evolution \citep{Reimers1975,Schroeder2005}. Old RGB stars lose approximately $0.2$\,\Msun\ by a magneto-hydrodynamic mechanism (Alfv\'en waves) during their ascent to the tip, and this amount of mass-loss has been observationally found to be fairly independent of mass and  metallicity \citep{Dupree2009,Gratton2010,Cranmer2011,McDonald2011ApJL,McDonald2011ApJS,Miglio2012,McDonald2015,Salaris2016,Tailo2021}. Red clump stars are likely to have lost similar amounts of mass as stars near the TRGB. Stellar encounters, in particular in globular clusters, could further enhance mass-loss rates \citep{Pasquato2014}.  At luminosities exceeding $\log{L/L_\odot} \approx 3$, RGB stars further become subject to dust-driven mass-loss \citep{Groenewegen2012}. Since the TRGB reaches very close to this cut-off luminosity, it remains an open question whether dust-driven mass loss could be relevant, perhaps especially for cooler, more metal-rich specimen. As pointed out by \citet{Cassisi2017}, there is not yet consensus on how to incorporate these effects in stellar evolution models, which typically assume the much weaker \citet{Reimers1975} mass-loss rates, or in population synthesis codes. Thus, we can currently only speculate about how strongly the material lost could redden or attenuate RGB stars.

Given these considerations, we expect localized extinction to be small, albeit non-zero, and compute \kcorr s for a range of small range of $\mathrm{E(B-V)^{host}} < 0.05$\,mag values to compute $A^{\mathrm{host}}$ to assess the combined effects. Unless otherwise specified, values of \Kext\ discussed in the following assume a typical host reddening of $\mathrm{E(B-V)^{host}}=0.01$\,mag due to the CGM alone. Any potential extinction by circumstellar environments (due to mass-loss) has not yet been sufficiently quantified and remains a subject for future research.

\section{Results \label{sec:results}}

We computed a grid of \kcorr s for a wide parameter range covering \teff\ between $3500$ and $6000$\,K; \logg\ between $2.0$ and $0.0$; [Fe/H] between $-2.0$ and $0.5$; redshift $z$ between $0.005$ and $0.030]$, and host reddening $\mathrm{E(B-V)^{host}}$ between $0.0$ and $0.5$\,mag in the {\it HST/WFC3} filters F555W, F814W, and F160W, 2MASS J \& $K_s$, and {\it JWST/NIRCAM} filters F115W, F150W, F200W, and F277W. As explained above, calibration and foreground extinction are set to zero for these single-filter calculations. Table\,\ref{app:tabgrid} illustrates a small subset of the grid. Table\,\ref{app:WesenCepheid} lists optical-NIR Wesenheit \kcorr s, RLB corrections, and the Wesenheit slope correction for Cepheids and $\mathrm{E(B-V)^{cal}}=0.4$\,mag, $\mathrm{E(B-V)^{fg}}=0.0023$\,mag, and a range of $\mathrm{E(B-V)^{host}}$ values. A separate grid was computed for stars near the TRGB for a smaller and denser range in $\mathrm{E(B-V)^{host}}$ between $0.000$ and $0.050$\,mag and including {\it HST/ACS} filters F606W and F814W, cf. Tab.\,\ref{app:TRGBebv}. Tables\,\ref{app:tabgrid}, \ref{app:WesenCepheid}, and \ref{app:TRGBebv} are made available in their entirety in the electronic appendix of the journal and via the CDS\footnote{\url{https://cds.u-strasbg.fr/}}.

\begin{figure*}
\centering
\includegraphics{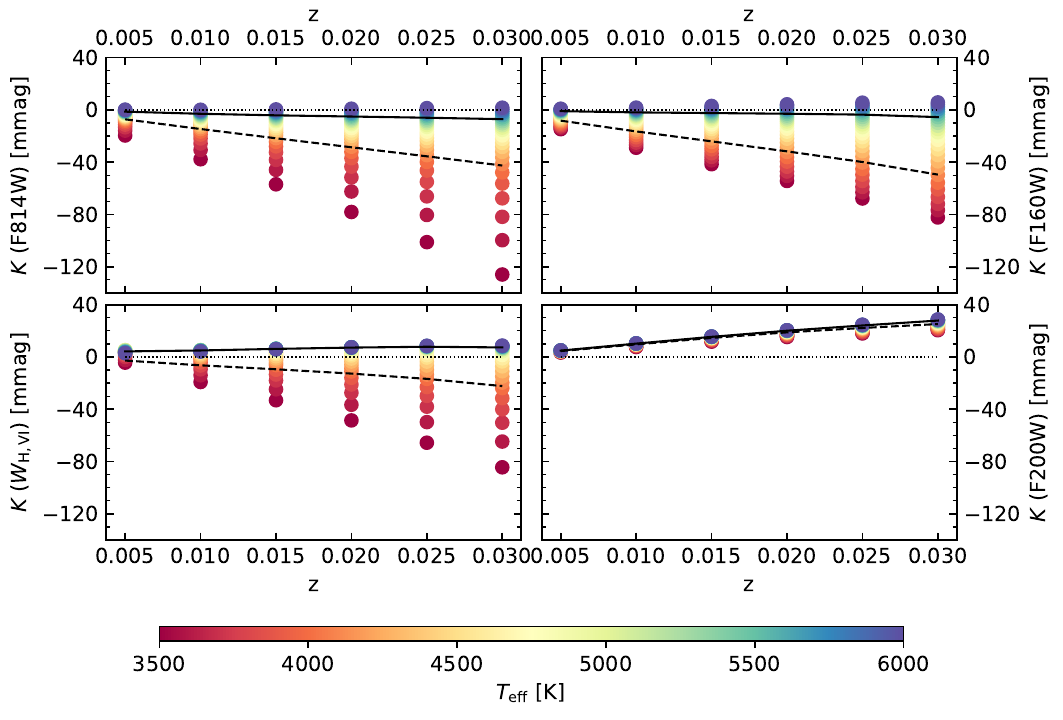}
\caption{\kcorr s as a function of redshift $z$ and \teff\ for different filters of interest for the simplified case without extinction. Circles show results for \teff$\in [3500,6000]$\,K (from red to blue circles, cf. below color bar), with fixed \logg$=1.5$ and [Fe/H]=$0.0$. The fiducial 10\,d MW Cepheid analog is shown by thick solid lines. Results for the fiducial TRGB star ([Fe/H]$=-1.75$) are shown as dashed thick black lines.\label{fig:ATLAS9K}}
\end{figure*}

\begin{figure}
\centering
\includegraphics{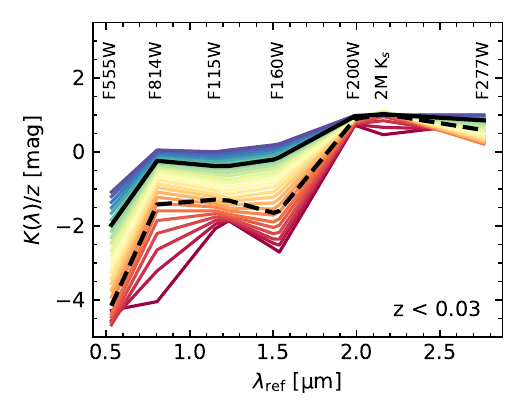}
\caption{\kcorr s divided by redshift $z$ as a function of filter reference wavelength for stars of different temperature (same scale as in Fig.\,\ref{fig:ATLAS9K}) for the simplified case without extinction. The fiducial 10\,d MW Cepheid analog is shown as a thick black solid line, and the fiducial TRGB star as a thick dashed black line. \kcorr s in the {\it JWST/NIRCAM} F200W and 2MASS $K_s$ filters are particularly insensitive to temperature differences. \kcorr s become increasingly positive and less filter-sensitive with wavelength. \label{fig:ATLAS_SpTypPassband}}
\end{figure}

Figure\,\ref{fig:ATLAS9K} illustrates the results of this grid computation for {\it HST} passbands ($\Lambda$) F814W, F160W, the NIR Wesenheit function, and the {\it JWST/NIRCAM} F200W filter for models with \logg$=1.5$, Solar iron abundance, and neglecting reddening. The run of $K(\Lambda,z)$ is smooth and monotonic for each temperature, though not necessarily linear. $K(\Lambda,z)$ becomes increasingly positive with wavelength, changing sign from negative to positive near $2\,\mu$m for most models, cf. Fig.\,\ref{fig:ATLAS_SpTypPassband}. $K(\Lambda,z)$ depends less and less on \teff\ with increasing wavelength, leading to large scatter in F814W and rather consistent values in F200W. At F277W, the spread in $K(\Lambda,z)$ is again larger than in F200W. At first sight, the NIR Wesenheit function mainly resembles the F160W variation. However, closer inspection shows that hotter \teff\ results are compressed together while the difference with the coolest stars is close to that seen in F160W, implying a significant non-linear dependence on \teff\ for cooler stars. Additional detail for Cepheids and TRGB stars is provided in the following subsections.

\subsection{Classical Cepheids\label{res:Cepheids}}

This subsection presents \kcorr s for Cepheids in two steps. Section\,\ref{res:10dCep} begins with a description of the fiducial $10$\,d MW Cepheid (cf. Sec.\,\ref{data:Cepheids}) to guide further application to \Ho\ measurements. In turn, Sec.\,\ref{res:tempCeph} discusses various additional effects, such as varying \teff\ due to pulsations, the intrinsic width of the IS, and the average IS position as a function of chemical composition. 

\subsubsection{The fiducial 10\,d MW Cepheid\label{res:10dCep}}

\begin{figure}
\centering
\includegraphics{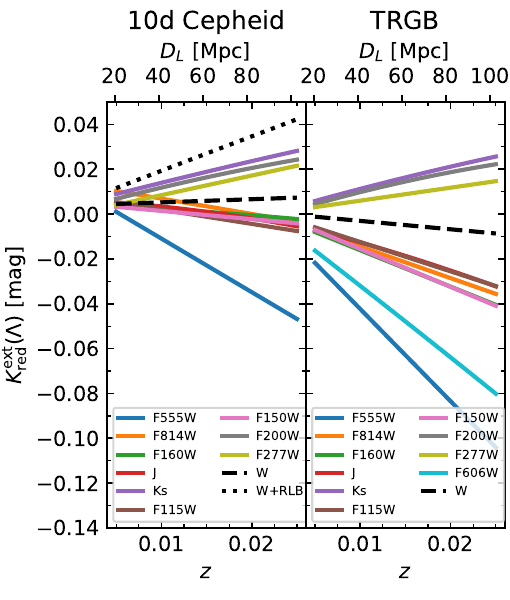}
\caption{\label{fig:Kofz_Cep+TRGB_allfilt}Illustration of extinction-dependent reduced \kcorr s as a function of redshift for a $10\,$d Cepheid listed in Tab.\,\ref{tab:Cep K(z) filters} and stars near the TRGB listed in Tab.\,\ref{tab:TRGB K(z) filters}. The results shown assume host reddening of $0.20$\,mag and $0.00$\,mag for Cepheids and TRGB stars, respectively. For Cepheids, \Kred\ includes a correction for extinction based on apparent color excess.
The NIR Wesenheit function and $K_s-$band exhibit the weakest $z-$dependence for Cepheids and remain generally very close to $0$. 
For Cepheids, the run of $K^{\mathrm{ext}}_{\mathrm{red}}(z)$ is difficult to discern due to overlapping lines for filters with wavelengths between F814W and F160W. TRGB \kcorr s overlap for filters F150W and F160W (green \& light pink), as well as for F115W and 2MASS $J-$band (brown \& red). }
The top x-axis shows luminosity distance $D_L$ (outer tick marks) in addition to redshift (inner ticks). The dotted black line shows the combination of \Kred\ and RLB (Eq.\,\ref{eq:deltaMuRel}) for the Wesenheit function, where \Kred\ and RLB add with the same positive sign. \Kw\ for TRGB stars and Cepheids have opposite sign due to the intrinsic color difference of these types of stars. 
\end{figure}

\begin{table}
\caption{\label{tab:Cep K(z) filters}Coefficients for $K-$corrections as a function of redshift for $10$\,d Cepheids}
\centering
\begin{tabular}{lccc}
$\Lambda$ & a & b & c \\
\hline
\multicolumn{4}{c}{{\it HST/WFC3}}\\
\hline
F555W  &  0.0130 & -2.3909 & \\
F814W  &  0.0137 & -0.6839 & \\
F160W  &  0.0044 & -0.1708  & -3.7457 \\
$W_{\mathrm{H,VI}}$ &  0.0038 &  0.1416 & \\
\hline
\multicolumn{4}{c}{2MASS}\\
\hline
$J$       &  0.0065 & -0.1980  & -11.1874 \\
$K_s$     &  0.0037 &  1.0587  & -3.1459 \\
\hline
\multicolumn{4}{c}{{\it JWST/NIRCAM}}\\
\hline
F115W  &  0.0084 & -0.6372 & \\
F150W  &  0.0051 & -0.3688 & \\
F200W  &  0.0013 &  1.1349  & -8.6439 \\
F277W  & -0.0003 &  0.8735 & \\
\hline
\end{tabular}
\tablefoot{$K-$corrections for 10\,d Cepheids are computed using $K^{\mathrm{ext}}_{\mathrm{red}}(\Lambda,z) = a + b\cdot z\, [ + c\cdot z^2 ]$\,mag. Coefficients are limited to $z< 0.03$ for linear relations and to $0.005 < z < 0.03$ for quadratic ones. Quadratic relations were adopted if non-linear behavior is readily apparent from inspecting the computed values of \Kred, which include host reddening equivalent to $\mathrm{E(B-V)^{host}}=0.20$\,mag that is corrected using apparent color excess.}
\end{table}

Figure\,\ref{fig:ATLAS9K} immediately shows that $K$ remains rather close to zero for F814W, F160W, and the Wesenheit function, and moves to increasingly positive values for the longer-wavelength F200W filter for the fiducial 10\,d Cepheid. To obtain a closer appreciation of the dependence of $K$ on redshift and filter, we computed \kcorr s for the fiducial Cepheid model and fitted linear or quadratic relations to \Kredz\ for each filter. We note that these values of \Kred\ are based on an assumed host reddening of $\mathrm{E(B-V)^{host}=0.20}$\,mag and a correction based on the apparent color excess (cf. Eq.\,\ref{eq:redKext}). Quadratic relations were adopted where a clear deviation from linear was visible by eye.
The fitted relations for \Kredz\ are listed in Tab.\,\ref{tab:Cep K(z) filters} and illustrated in the left panel of Fig.\,\ref{fig:Kofz_Cep+TRGB_allfilt}. 

Several aspects are readily apparent in Fig.\,\ref{fig:Kofz_Cep+TRGB_allfilt}: 1) \Kred\ can deviate $0$ at low $z$ due to imperfect extinction corrections based on apparent color excess (most apparent in F814W); 2) the sign of $K$ changes from negative to positive at wavelengths slightly longer than F160W ($H-$band); 3) $K$ for the Wesenheit function is very close to $0$ and does not strongly depend on $z$; 4) $K$ exceeds $>1\%$ in distance for objects farther than 100\,Mpc in filters F555W, $K_s$, F200W, and F277W. 5) The F555W distance error exceeds $1\%$ at $\sim 50$\,Mpc. 

The insensitivity of the NIR Wesenheit function to \kcorr s is quite remarkable. 
Changes in the assumed value of $R_V$ used for computation of the Wesenheit slope do not significantly change this picture. Applying the redshift-correction to the Wesenheit slope (Eq.\,\ref{eq:Rofz}) also does not visibly affect the run of $K\left(W_{\mathrm{H,VI}},z\right)$ on the scale shown in Fig.\,\ref{fig:Kofz_Cep+TRGB_allfilt}.

Figure\,\ref{fig:Kofz_Cep+TRGB_allfilt} also hints at the interesting possibility of directly measuring \Kred\ at the longer redshift ranges by comparing distances to the same SN-host galaxies observed in filters with oppositely signed \kcorr s. While this will be extremely challenging for Cepheids, TRGB stars (right panel, Sec.\,\ref{res:TRGB}) may exhibit a difference of up to $\Delta \mu \approx 0.06$\,mag between the {\it JWST/NIRCAM} filters F200W and F150W. Directly detecting such a systematic chromatic dependence of measured distance modulus on $z$ would provide a direct confirmation of the relativistic effects associated with cosmic expansion.

To compute the full relativistic distance bias, Eq.\,\ref{eq:deltaMuRel} adds to \Kred\  contributions by RLB and the Wesenheit slope-dependence on $z$. Figure\,\ref{fig:Kofz_Cep+TRGB_allfilt} illustrates the impact RLB for the example of the Wesenheit function $K\left(W_{\mathrm{H,VI}},z\right)$. RLB depends on the photometric filter due to the LL slope dependence on wavelength (steeper at longer wavelength). Hence, the offset applicable to other photometric filters differs in detail, but not in direction, since the LL slope is always negative. Thus, for all photometric filters, RLB would bring \Kred\ in F555W much closer to $0$, whereas it would tend to exacerbate the positively signed values of \Kred\ for longer-wavelength filters.

\subsubsection{Temperature-dependence: Leavitt law slope and phase-dependence\label{res:tempCeph}}

The temperature-dependence of $K$ (Fig.\,\ref{fig:ATLAS9K}) impacts Cepheids in two ways. Firstly, the \teff-dependence on \logP\ illustrated in Fig.\,\ref{fig:PCrelation} leads to a \logP-dependence of $K$, which leads to a small LL slope-dependence on redshift. This effect correlates with the selection bias of preferentially detecting longer-period Cepheid in more distant galaxies due to their greater intrinsic luminosity. Secondly, Cepheid pulsations modulate \teff\ in excess of $1000$\,K between phases of minimum and maximum light, leading to phase-dependent \kcorr s.

\begin{table}
\caption{\label{tab:KlogP}Filter-dependent $K-$corrections for fiducial 10\,d Cepheids}.
\centering
\begin{tabular}{@{}llrrr@{}}
\hline
$\Lambda$ & redshift & distance & slope & intercept \\
 & & (Mpc) & (mmag/\logP) & (mmag) \\
 \hline
F555W               & 0.0019 & 7.6 & -2.84 & -1.74  \\
F814W               & 0.0019 & 7.6 & -1.02 & -0.17  \\
F160W               & 0.0019 & 7.6 & -1.18 &  1.00  \\
$W_{\mathrm{H,VI}}$ & 0.0019 & 7.6 &  3.48 &  0.51  \\
 \hline
F555W               & 0.0056 & 23  &  -8.65 & -5.47  \\
F814W               & 0.0056 & 23  &  -3.11 & -0.91 \\
F160W               & 0.0056 & 23  &  -3.53 &  1.93 \\
$W_{\mathrm{H,VI}}$ & 0.0056 & 23  &   2.68 &  1.74 \\ 
\hline 
F555W               & 0.0098 & 40  & -15.16 & -9.48 \\
F814W               & 0.0098 & 40  &  -5.47 & -1.79 \\
F160W               & 0.0098 & 40  &  -6.04 &  3.19 \\
$W_{\mathrm{H,VI}}$ & 0.0098 & 40  &   1.89 &  3.25 \\
\hline
F555W               & 0.0172 & 70  & -26.85 & -15.67  \\
F814W               & 0.0172 & 70  &  -9.40 &  -2.82  \\
F160W               & 0.0172 & 70  & -10.10 &   5.69 \\
$W_{\mathrm{H,VI}}$ & 0.0172 & 70  &   1.07 &   5.96 \\
\hline
F555W               & 0.0245 & 100 & -38.66 & -20.51 \\
F814W               & 0.0245 & 100 & -12.73 &  -4.02 \\
F160W               & 0.0245 & 100 & -14.38 &   8.28 \\
$W_{\mathrm{H,VI}}$ & 0.0245 & 100 &   0.31 &   8.05 \\
\hline
\end{tabular}
\tablefoot{Coefficients for extinction-dependent reduced $K-$corrections, $K^{\mathrm{ext}}_{\mathrm{red}}(\log{P},z) = a \cdot \log{P} + b$, with slope $a$ and intercept $b$, computed assuming $\mathrm{E(B-V)^{host}}=0.20$\,mag.\label{tab:KlogP}}
\end{table}

\begin{figure}
\centering
\includegraphics{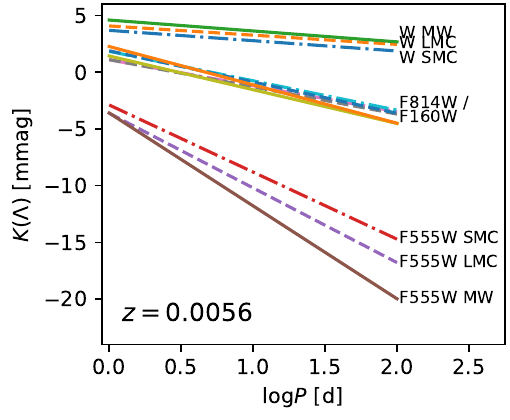}
\caption{\kcorr s for Cepheids as function of \logP\ for different metallicities, computed assuming a redshift difference of $0.0056$ ($\sim 23$\,Mpc) for several filters and assuming no extinction. Line styles identify the adopted metallicity of the MW (solid lines), the LMC (dashed lines), and the SMC (dash-dotted lines).\label{fig:CephMet}}
\end{figure}

We computed $K(\log{P},z) = m \cdot \log{P} + c$ at several redshifts and for several filters $\Lambda$ by converting input values of \logP\ to \teff\ and \logg\ using Eqs.\,\ref{eq:Teff_from_logP} and \ref{eq:logg_from_logP}. Figure\,\ref{fig:CephMet} illustrates the results for SN-host galaxies at $23$\,Mpc, and Tab.\,\ref{tab:KlogP} lists the coefficients determined for a range of different redshifts from NGC\,4258 to the Coma cluster. Interestingly, the slope for the Wesenheit function is close to $0$ over the redshift range where it is reasonable to hope for Cepheids to be observed in the foreseeable future. 
For individual filters this is not necessarily the case, however. In F160W, the slope effect leads to a difference of $\sim 0.014$\,mag per \logP\ between nearby Cepheids and Cepheids in the Coma cluster.

\label{res:metalCeph}

Figure\,\ref{fig:CephMet} also illustrates the impact of chemical composition on Cepheid \kcorr s, since the center position of the instability strip moves to hotter temperatures at lower metallicity (Sec.\,\ref{data:Cepheids}). It shows \kcorr s computed at $z = 0.0056$ ($\sim 23$\,Mpc) for the NIR Wesenheit function and its constituent filters computed for a range of input \logP\ values and iron abundances matching the  metallicity of the Sun, the LMC, and the SMC Cepheid populations: $[\mathrm{Fe/H}]=0.0$, $-0.3$, and $-0.7$. As could be expected, the \logP-dependence of \Kext\ is only slightly modified by metallicity. The largest effect is seen in F555W, which coincides nearly with the peak of the Cepheid SED. We conclude that chemical composition has no significant impact on \kcorr s of Cepheids.

\begin{figure}
\centering
\includegraphics{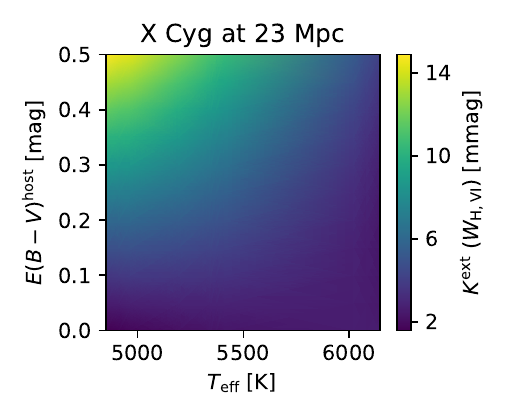}
\caption{\label{fig:XCyg_phase_ebv}Wesenheit \kcorr s for an X Cyg-like Cepheid (16.4\,d) at 23 Mpc as a function of phase (by proxy of \teff) and assuming different amounts of host reddening. The most significant phase variation is seen for the highest amount of reddening, where \Kw\ may differ by as much as $0.01$\,mag among phases of extreme \teff\ values.}
\end{figure}

Figure\,\ref{fig:XCyg_phase_ebv} illustrates phase- and host reddening-dependence for a long-period Cepheid-analog, X~Cyg ($P = 16.4$\,d) with \teff$=5267$\,K and a temperature amplitude of $\sim 1300$\,K \citep{Kovtyukh2005}, observed at $z = 0.0056$ ($23$\,Mpc). In low-reddening environments, phase variations of $K$ are generally at the single-digit mmag level and can be safely ignored for distance measurements determined on cycle-averages. 
The phase-dependence becomes greatest at hotter temperatures and for high host reddening, but remains limited to $\lesssim 0.01$\,mag between phases of extreme \teff\ differences. Thus, temperature variations due to pulsations do not significantly modify \Kw.

\subsection{\kcorr s for TRGB distances\label{res:TRGB}}
The right panel of Fig.\,\ref{fig:Kofz_Cep+TRGB_allfilt} illustrates the dependence of \Klam\ for TRGB stars observed in various filters assuming no host reddening, as tabulated in Tab.\,\ref{tab:TRGB K(z) filters}. The transition from negative to positive values of $K$ happens at the same reference wavelength as for Cepheids; roughly at $2\,\mu$m (F200W). For filters where $K$ is positive, the pattern resembles that seen in the left panel for Cepheids, with marginally larger spread. For filters where $K$ is negative, however, the spread grows more rapidly as a function of $z$ than for Cepheids. We find a $\sim 0.06$\,mag difference in \Klam\ between filters F150W and F200W at Coma cluster distances, which could allow for a future observational test of the predicted \kcorr s using \jwst. The optical filters F555W and F606W require the  largest negative $K-$corrections, which already reach or exceed $\sim 1\%$ at 20\,Mpc.

For low-metallicity ([Fe/H]=-1.75) TRGB stars, differences in $K(\mathrm{F814W})$ within the temperature range \teff$= 4200 \pm 100$\,K, computed as hotter minus cooler star, are at the level of $-0.6$\,mmag ($-3$\,mmag) at $20$\,Mpc ($100$\,Mpc). Varying instead metallicity between [Fe/H]$=-1.75$ and $-0.75$ yields similar differences, computed as more metal-poor minus more metal-rich star, of $-0.7$\,mmag and $-4.4$\,mmag at 20\,Mpc and 100\,Mpc, respectively.

The impact of host extinction (cf. Sec.\,\ref{data:TRGB}) on TRGB distances was considered by computing a smaller, separate grid for temperatures near the fiducial TRGB \teff, \logg, low metallicity, and a finely sampled range of small $\mathrm{E(B-V)^{host}}$ values. These results are included as Tab.\,\ref{app:TRGBebv} in the online appendix. Even small amounts of host reddening can cause \Ki\ to change sign from negative to positive. For example, $K^{\mathrm{ext}}(\mathrm{F814W}) = -0.007$ ($-0.036$) and $+0.011$\,mag ($-0.018$) at $\sim 20$\,Mpc ($100$\,Mpc) for $\mathrm{E(B-V)^{host}}=0.0$ and $0.01$, respectively. 
At longer wavelengths, reddening has a much reduced effect, of course, leading to $K^{\mathrm{ext}}(\mathrm{F160W}) = -0.008$ ($-0.040$) and $-0.002$ ($-0.034$) mag, respectively. 

\begin{figure}
\includegraphics{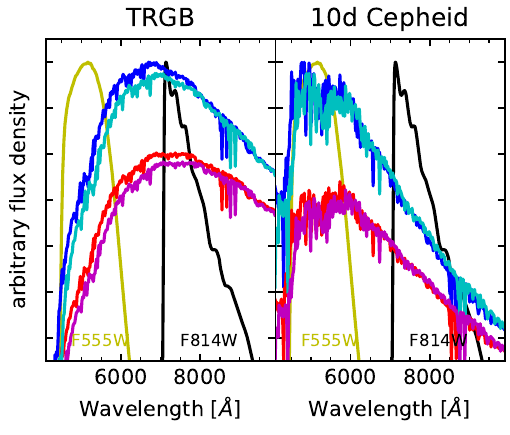}
\caption{\label{fig:SEDfilter}Spectral energy distribution incident on {\it HST/WFC3} filters F555W (yellow) and F814W (black) for a typical TRGB star (left) and a 10-day Cepheid (right). Unreddened spectra at rest are plotted in dark blue, the unreddened spectra at $z=0.025$ in cyan. Spectra reddened by $E(B-V)=0.10$\,mag are shown at rest (red) and at $z=0.025$ (magenta) for illustration purposes. The sensitivity of stars near the TRGB observed in F814W to even light reddening and small redshift is readily apparent and mirrors that of 10\,d Cepheids observed in F555W.}
\end{figure}

The sensitivity of \Ki\ to reddening and redshift can be understood by considering that SEDs of early K giants peak slightly blueward of the F814W filter, cf. Fig.\,\ref{fig:SEDfilter}. This leads to two effects: 1) the exponential nature of dust attenuation (Eq.\,\ref{eq:extinction}) significantly reduces the flux incident on the F814W filter even for a very low extinction values; 2) reddening progressively shifts the SED peak from blueward of the F814W filter to squarely within it, thereby changing the SED slope incident on the filter and causing sign reversal of $K$.

The Wesenheit function could help reduce sensitivity to such issues thanks to its very small \kcorr\ across the redshift range considered. TRGB distances based on the Wesenheit function would be insensitive to relativistic effects and reddening at the $1\%$ level even at 100\,Mpc. Even if no reddening is considered (as in right panel of Fig.\,\ref{fig:Kofz_Cep+TRGB_allfilt}), the Wesenheit \kcorr\ is smaller in absolute value than the one applicable to F277W. Further benefits of the Wesenheit function include reduced sensitivity to \teff\ and metallicity, that is, the parameters of the TRGB host populations.

\begin{table}
\caption{\label{tab:TRGB K(z) filters}Filter-dependent TRGB $K-$corrections as function of redshift}
\centering
\begin{tabular}{lccc}
$\Lambda$ & a & b & c \\
\hline
\multicolumn{4}{c}{{\it HST/WFC3}}\\
\hline
F555W  & -0.0012 & -4.1162 & \\
F814W  & -0.0004 & -1.4075 & \\
F160W  &  0.0001 & -1.6241 & \\
\hline
\multicolumn{4}{c}{{\it HST/ACS}}\\
\hline
F606W &  -0.0000 & -3.2065 & \\
\hline
$W_{\mathrm{H,VI}}$ &  0.0007 & -0.3744 & \\
\hline
\multicolumn{4}{c}{2MASS}\\
\hline
$J$    & -0.0021 & -0.9549  & -10.1038 \\
$K_s$  &  0.0001 &  1.1706  & -5.9837 \\
\hline
\multicolumn{4}{c}{{\it JWST/NIRCAM}}\\
\hline
F115W  &  0.0006 & -1.3100 & \\
F150W  &  0.0013 & -1.6901 & \\
F200W  & -0.0015 &  1.2513  & -12.0468 \\
F277W  &  0.0002 &  0.5791 & \\
\hline
\end{tabular}
\tablefoot{Linear or quadratic relations for $K(\Lambda,z) = a + b\cdot z [ + c\cdot z^2 ]$\,mag applicable to TRGB stars for $z< 0.03$ for linear relations and for $0.005 < z < 0.03$ for quadratic ones. Quadratic relations were adopted if non-linear behavior is discernible from inspecting the computed values of $K$, which assume no host reddening ($\mathrm{E(B-V)^{host}}=0.0$\,mag).}
\end{table}

\subsection{Comparing $\Delta \mu^{\mathrm{rel}}$ for the TRGB and Cepheids\label{sec:reddening}}

The most recent Cepheids-based \Ho\ value from \citet[\Ho$=73.2\pm1.3$\Hunit]{Riess2021} and the TRGB-based value by \citet[\Ho$=69.8\pm1.7$\Hunit]{Freedman2021} differ by $\Delta \mu \approx 5 \log{(73.2/69.8)} = 0.103$\,mag, and the origin of this $1.6\sigma$ difference is much discussed \citep[e.g.][]{Freedman2021,Anand2021edd}. Reducing this difference to $1.0\sigma$ would require a shift of $\Delta \mu_{1\sigma} = 0.065$\,mag.  
Following the above general discussion of \kcorr s for Cepheids and TRGB stars, this subsection seeks to understand whether the investigated mechanisms can reconcile the two measurements at $1\sigma$. Additional considerations for the recent TRGB methodology employed by the Cosmicflows team \citep{Tully2013} are presented in Sect.\,\ref{sec:EDD} below.

To this end, we computed the combined relativistic distance correction (Eq.\,\ref{eq:deltaMuRel}) at the current average SN-host redshifts. 
For Cepheids, we used temperatures representative of pulsation periods $P=5$\,d to $60$\,d as per Eq.\,\ref{eq:Teff_from_logP}, while for TRGB stars we investigated a broad range $\pm 200$\,K around the fiducial temperature of $4200$\,K. For Cepheids, we considered wider range of host reddenings ($\mathrm{E(B-V)^{host}}$ between $0.0$ and $0.5$\,mag) than for TRGB stars ($\mathrm{E(B-V)^{host}}$ between $0.0$ and $0.05$). We considered measurements based on the Wesenheit function for Cepheids, whereas we computed \Ki\ for TRGB stars, i.e., the extinction-dependent \kcorr\ without applying a correction for host extinction.

\begin{figure*}
\centering
\includegraphics{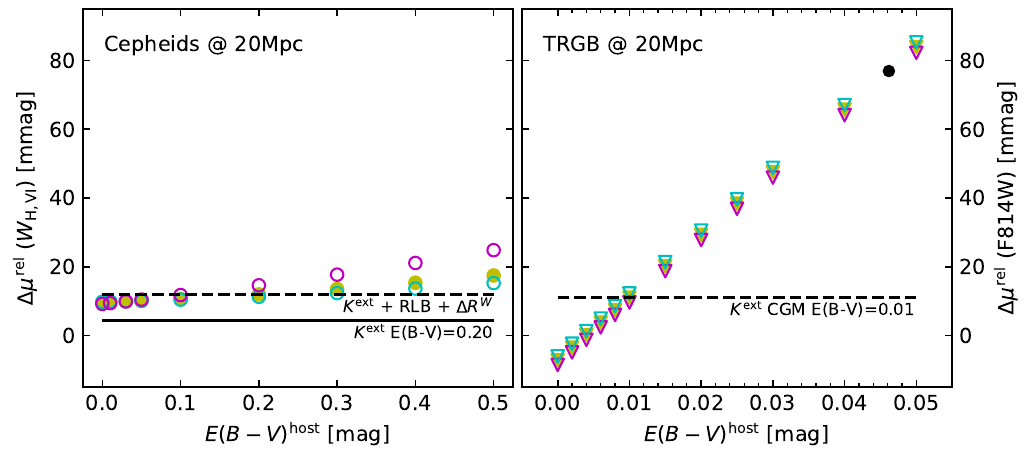}
\caption{\label{fig:CepTRGBEBV}Distance modulus bias at $20$\,Mpc as function of host reddening for Cepheids observed using the optical-NIR Wesenheit function (left) and the TRGB calibrated in the F814W filter (right). Symbols show stars of different temperature, for Cepheids corresponding to pulsation periods of 5, 10, and 60\,d (cyan, yellow, magenta circles, respectively) and for TRGB stars for fiducial \teff\ of 4100, 4200, and 4300\,K (magenta, yellow, cyan downward triangles). The solid horizontal line for Cepheids indicates the \kcorr\ alone for host reddening corresponding to E(B-V)=0.20\,mag. The dashed horizontal lines illustrate the preferred corrections; for Cepheids, this includes corrections for RLB and the redshifted reddening law, for TRGB stars a halo extinction of E(B-V)=0.01\,mag. The solid black circle in the right panel illustrates the amount of host reddening at which Cepheid and TRGB-based \Ho\ measurements would become consistent to within their joint $1\sigma$ uncertainty. The ordinate of this point is the sum of the $0.012$\,mag correction applicable to Cepheids and the current $0.065$\,mag shift required to render both (uncorrected) \Ho\ measurements consistent within $1\,\sigma$.} 
\end{figure*}

Figure\,\ref{fig:CepTRGBEBV} illustrates both results at $z = 0.005$, which is close to the average SN-host distances of approximately $20$\,Mpc \citep{Riess2016,Freedman2021}. The dependence of \Kw\ on host extinction is very minor for the fiducial Cepheid, and only slightly larger for the cooler, longer-period Cepheids, cf. also Sec.\,\ref{res:tempCeph}. Assuming a typical host extinction of $E(B-V)=0.20$ and accounting for RLB as well as the redshifted reddening law yields a distance bias correction of $\Delta \mu^{\mathrm{rel}} \approx 0.012$\,mag, of which $0.007$\,mag are due to RLB, which quickly becomes the dominant relativistic effect at greater redshifts.

For TRGB stars, $K^{\mathrm{ext}}(\mathrm{F814W})$ exhibits a rather steep dependence on E(B-V), which could be reduced by applying reddening corrections, e.g. via the apparent color excess relative to a fiducial color (Eq.\,\ref{eq:redKext}). 
At $z=0.005$, $\Delta \mu^{\mathrm{rel}}$ exhibits a slope of approximately $0.018$\,mag per $0.01$\,mag $\mathrm{E(B-V)^{host}}$. As mentioned in Sec.\,\ref{data:TRGB}, studies of the CGM have measured CGM reddening on the order of $0.01$\,mag in resolved galaxies at impact parameters of up to a few tens of kpc where TRGB measurements are made \citep[e.g.][]{Menard2010,Peek2015}. Adopting the typical value of $\mathrm{E(B-V)^{host}} = 0.01$\,mag, we find $\Delta \mu^{\mathrm{rel}} \approx 0.011$\,mag, which is nearly identical to the correction applicable to Cepheids. 

Additional uncorrected localized extinction applicable to TRGB distances may arise due to circumstellar environments (CSEs), although the magnitude of this effect and its dependence on wavelength remain open questions at this time. Nevertheless, any amount of additional host reddening increases $\Delta \mu^{\mathrm{rel}}$ and reduces the difference in \Ho\ measured using Cepheid and TRGB calibrations of type-Ia supernovae. Hence, it is interesting to check whether two \Ho\ measurements could be reconciled to within a combined $1\sigma$ confidence interval by localized extinction, which would require a correction of $\Delta \mu^{\mathrm{rel}} = 0.065 + 0.012 = 0.077$\,mag, which would correspond to $\mathrm{E(B-V)^{host}} = 0.046$\,mag, cf. the filled black circle in the right panel of Figure\,\ref{fig:CepTRGBEBV}.

To test a possible contribution of CSEs to localized extinction, we considered the difference in the absolute F814W magnitude reported for the TRGB in the Magellanic Clouds \citep{Hoyt2021}, NGC\,4258 \citep{Jang2021}, and 46 MW globular clusters \citep{Cerny2020}. Since the reddening maps of the Magellanic Clouds \citep{Skowron2021} used by Hoyt are based on red clump stars, one may expect that these stars would have experienced similar mass loss as stars near the TRGB. On the other hand, the reddening corrections applied to NGC\,4258 \citep[all-sky reddening maps by][]{Schlegel1998,Schlafly2011} and the MW TRGB \citep[color-magnitude diagrams of main sequence cluster members compiled in][2010 edition]{Harris1996} should not be sensitive to CSEs. 

Hence, the difference between the averages of both calibration groups may be sensitive to CSE reddening, and we find $\Delta M_{\mathrm{F814W}} = \langle M_{\mathrm{F814W}} \rangle_{\mathrm{LMC,SMC}} - \langle M_{\mathrm{F814W}} \rangle_{\mathrm{N4258,MW}} = 0.005$\,mag. 
Although the sign of this difference is consistent with an unaccounted for extinction contribution among the TRGB measured in NGC\,4258 and the MW (the calibrated absolute magnitude is too bright if extinction is not adequately corrected), we note that the difference of $0.005$\,mag is much smaller than the $0.046 - 0.010 = 0.036$\,mag required to render Cepheid and TRGB \Ho\ values consistent to within $1\sigma$. Hence, our methodology cannot reconcile the reported disagreements between Cepheid and TRGB distances, unless CSEs of TRGB stars contribute additional localized reddening at the level of $0.04$\,mag. We stress that this simple consideration should not be considered a detection of reddening due to CSEs. Detailed research is required to assess the possible impact of stellar mass-loss and CSEs on TRGB distances.

Interestingly, relativistic corrections for F814W observations of the TRGB and the Wesenheit function of Cepheid stars increasingly differ at greater redshifts. At $z=0.025$, we find $\Delta \mu^{\mathrm{rel}}_{\mathrm{Cep}} = 0.008 + 0.035 + 0.003 = 0.046$\,mag (\Kext, RLB, and $\Delta R^W$ terms listed separately). Thus, RLB dominates over the effect of \kcorr s at distances of order 100\,Mpc.
For the F814W TRGB and host reddening of $\mathrm{E(B-V)^{host}}=0.01$\,mag, we find $\Delta \mu^{\mathrm{rel}}_{\mathrm{TRGB}}= - 0.018$\,mag. It remains to be seen whether future observations will be precise enough to unambiguously detect this small systematic difference of $0.064$\,mag between the two methods.

\begin{figure}
\centering
\includegraphics{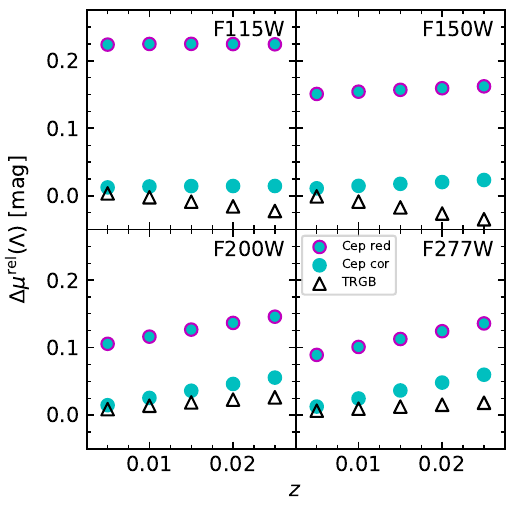}
\caption{\label{fig:Kred_JWST}Combined relativistic distance corrections for Cepheids (circles) and the TRGB (triangles) observed in four broad-band \jwst\ passbands as a function of redshift. Host reddenings of E(B-V)$=0.01$\,mag and $0.20$\,mag have been assumed for the TRGB and Cepheids, respectively. The filled cyan circles show results for Cepheids after applying corrections for apparent host reddening, whereas the cyan circles with magenta envelopes show Cepheids uncorrected for host extinction. For Cepheids, RLB has been included assuming slopes of $-3.2, -3.3, -3.4, -3.45$\,mag/dex for the F115W, F150W, F200W, and F277W passbands, respectively.}
\end{figure}

Figure\,\ref{fig:Kred_JWST} compares the combined relativistic distance corrections applicable to the TRGB and Cepheids for broadband \jwst\ passbands as a function of redshift and assuming host reddenings of $0.01$\,mag and $0.20$\,mag, respectively. For Cepheids, RLB corrections are also included using approximate LL slopes \citep[inspired by multi-wavelength results in][]{Madore2017}. The figure clearly illustrates the importance of accurate reddening corrections for Cepheids, even at NIR magnitudes, which requires accurate knowledge of fiducial colors. It also makes clear that relativistic corrections become increasingly important at greater redshift, in particular for Cepheids observed with the longer-wavelength filters where \Kred\ and $\Delta \mu^{\mathrm{RLB}}$ are both positive. For TRGB-based distances the correction remains just below the $1\%$ when observing in F277W even at 100\,Mpc, though it exceeds $\pm 1\%$ in the other three filters considered. The F277W's greatly reduced sensitivity to extinction and the smallness of \kcorr s in this filter are particularly useful for future TRGB observations with \jwst. For Cepheids, the Wesenheit function\hbox{---}in particular as a combination of \jwst\ filters to benefit from the improved spatial resolution compared to {\it HST/WFC3}\hbox{---}will likely remain important to avoid complications due to reddening, while RLB corrections are straightforward once the LL slope and redshift are known.

\subsection{\kcorr s for TRGB distances in the EDD \label{sec:EDD}}

The recent update of the Extragalactic Distance Database \citep[EDD]{Tully2009,Anand2021edd} contains TRGB distances to more than 500 nearby galaxies observed with {\it HST}. The same methodology (henceforth: the EDD TRGB method) was applied by \citet{Anand2021H0} to measure \Ho\ using a TRGB-calibrated distance ladder whose systematics differ as much as possible from the systematics of the TRGB distance ladder by \citet[i.e., the CCHP TRGB method]{Freedman2021}. Key differences among the systematics underlying the EDD and CCHP methods include\footnote{
While some of these differences, such as photometric homogeneity, are clear advantages of the EDD methodology over that adopted by the CCHP, the effect of other differences remains contentious, such as the benefit of relying on a single anchor galaxy with homogeneous photometry (EDD) versus averaging over multiple hosts with inhomogeneous photometry (CCHP), or the equivalence of TRGB populations in anchor and SN-host galaxies. Certainly, the detailed analysis of these differences and their relative importance for measuring \Ho is beyond the scope of this article. Interested readers are referred to the original EDD and CCHP articles for further detail.}
: a) the EDD's use of a maximum likelihood fit to a combined RGB/AGB luminosity function \citep{Makarov2006} instead of the CCHP's use of a Sobel filter for edge detection; b) the EDD's exclusive reliance on {\it HST/ACS} F606W observations obtained after the 2009 {\it HST} electronics upgrade; c) the EDD's absolute calibration using a different field of NGC\,4258 and its use as the sole distance ladder anchor; d) the EDD's correction for population variance based on a fiducial color calibration \citep{Rizzi2007}; e) the EDD's homogeneous treatment of reddening corrections based almost exclusively\footnote{the sole exception being the SN-host galaxy NGC\,5643 whose low Galactic latitude renders all-sky reddening maps unreliable} on all-sky maps for both the anchor galaxy and the SN-hosts. The following considers the interplay between color corrections and redshift for the EDD TRGB method; further details on the EDD methodology can be found in \citet{Makarov2006,Rizzi2007,Jacobs2009,Wu2014TRGB,Anand2021edd}. 

The EDD method employs the \citet{Rizzi2007} calibration to adapt the absolute F814W magnitude of the TRGB based on observed, de-reddened F606W - F814W color. This color calibration was derived using 5 nearby galaxies covering a large range in metallicity and in the {\it HST/ACS} F814W system yields $M_{\rm{F814W}} = -4.06 + 0.2\cdot \left( (\mathrm{F606W}-\mathrm{F814W} \right) - 1.23)$. According to Tab.\,2 in \citet{Anand2021H0}, there is a $0.049$\,mag difference between the TRGB color in NGC\,4258 and the average SN-host TRGB, with SN-hosts being redder. Hence, the average absolute TRGB F814W magnitude in SN-hosts is $0.01$\,mag fainter than in NGC\,4258. However, this correction does not account for the redshift difference\footnote{Computed using the difference in heliocentric redshifts tabulated by NED, with the exception of NGC\,1441, where we used the average of the Fornax cluster from \citet{Madore1999}} between the SN-host average and NGC\,4258 of $\Delta z^{\mathrm{SN}}_{\mathrm{cal}} = 0.00403 - 0.00149 = 0.00254$, which shifts the apparent color of the more distant TRGB stars redward. Using the fiducial TRGB model (cf Sec.\,\ref{data:TRGB}), we estimate $\Delta z^{\mathrm{SN}}_{\mathrm{cal}} = 0.00254$ to lead to an apparent color change of $\sim 0.005$\,mag, which implies a minute overcorrection by $\Delta \mu^{\mathrm{EDD}} = 0.001$\,mag based on the \citet{Rizzi2007} calibration. These color-redshift corrections increase linearly with redshift and will need to be included when measuring TRGB distances using the EDD methodology at increasing distances with \jwst.

Using our fiducial TRGB star, we find $K^{\rm{ext}}(\mathrm{F814W}) = -0.004$\,mag applicable to the EDD \Ho\ measurement. Complications due to additional host reddening are minimal because the sole anchor galaxy, NGC\,4258, is subject to the same reddening systematics as the EDD SN-hosts (only foreground reddening is corrected using all-sky dust maps). Adopting unaccounted for CGM reddening of $\mathrm{E(B-V)}=0.01$\,mag in both NGC\,4258 and the SN-hosts yields a difference in $K^{\rm{ext}}(\mathrm{F814W})$ of $\ll 0.001$\,mag and can be neglected.

\subsection{Implications for the Hubble constant\label{res:H0}}

Section\,\ref{sec:reddening} reported similar combined relativistic corrections for TRGB and Cepheid distances for an average SN-host distance of 20\,Mpc. Here, we apply this correction to Eq.\,\ref{eq:H0true} to determine the impact of relativistic corrections on published \Ho\ measurements.

For the average redshift of SN-host galaxies used by \citet{Riess2016} and the latest Cepheid-based \Ho\ measurement reported by \citet[\Ho$=73.2\pm1.3$\Hunit]{Riess2021}, the total bias $\Delta\mu^{\mathrm{rel}}_{\mathrm{Cep}}(0.0056) = 0.0133 \pm 0.0015$\,mag\footnote{the uncertainty on the correction is based on varying the amount of host reddening by $\pm 0.1$\,mag around the adopted typical $0.20$\,mag.}. This implies a previous systematic underestimate of \Ho\ by approximately $0.6\%$ due to the missing \kcorr, RLB correction, and the change in Wesenheit slope as a function of redshift. This translates to an \emph{increase} of \Ho\ by $+0.45 \pm 0.05$\Hunit\ to a final bias-corrected value of \Ho$=73.65 \pm 1.3$\Hunit, where the uncertainty of the correction does not noticeably contribute to the total error budget. The increased \Ho\ value also enhances the discord with respect to the early-Universe \Ho\ value reported by the \citet[\Ho$=67.4\pm0.5$\Hunit]{Planck2020H0} from $4.2\sigma$ to $4.4\sigma$. Additional uncertainty of the correction related to source color\hbox{---}e.g. longer period Cepheids being preferentially detected at greater distances\hbox{---}tends to further increase the numerical value of the correction, thus further increasing \Ho. We therefore consider this correction based on the fiducial 10\,d Cepheid to be a conservative estimate.

For the CCHP TRGB methodology, we obtained a correction of $\Delta \mu^{\mathrm{rel}}(\mathrm{F814W}) = 0.011 \pm 0.008$\,mag for an SN-host galaxy redshift of $0.0047$ \citep{Freedman2019}. The  stated uncertainty corresponds to results spanning the range of $\mathrm{E(B-V)^{host}}$ between $0.005$ and $0.015$\,mag. Applying this $0.5\%$ shift to the value of $H_0 = 69.8 \pm 1.7$\Hunit recently reported by \citet{Freedman2021} increases \Ho\ by $0.36 \pm 0.26$\,\Hunit to $H_0 = 70.2 \pm 1.7$\,\Hunit, which differs by $1.6\sigma$ from \Ho\ reported by the \citet{Planck2020H0}.

For the EDD TRGB methodology, we find a combined distance modulus correction of $K^{\rm{ext}}(\mathrm{F814W}) + \Delta \mu^{\mathrm{EDD}} = -0.004 + 0.001 = -0.003$\,mag, which translates to a reduction by a mere $-0.15\%$ and a reduction of \Ho\ by $0.1$\,\Hunit\ to $H_0 = 71.4 \pm 1.8$\,\Hunit. As pointed out by \citet{Anand2021H0}, this value is fully consistent with the Cepheids-based \Ho\ measurement of $72.0 \pm 1.9$\,\Hunit\ that assumes the same distance to NGC\,4258 for both types of standard candles \citep{Reid2019}. The different systematics concerning host reddening lead to oppositely signed corrections for the CCHP and EDD \Ho\ values and render them consistent to within $1\sigma$.

\section{Discussion\label{sec:disc}}

\subsection{Theoretical vs. empirical SEDs \label{disc:empiricalSED}\label{disc:models}}

We initially sought to avoid using stellar atmosphere models, and to instead employ empirical SEDs from stellar flux atlases, notably the Pickles stellar flux library \citep{Pickles1998} and the second data release of the XSHOOTER spectral library \citep[XSL2]{XSL2}. Unfortunately, the precision afforded by these data sets proved insufficient for our purposes because of several issues including flux calibration, telluric features that do not apply to space-based photometry, reddening of observed SEDs, incomplete or inhomogeneous wavelength coverage (Pickles), cross-calibration across different spectral orders (XSL2) or even among observations taken from different sites at different times (Pickles), time-dependent variable star SEDs, limited information on spectral types and \teff, reduced ability to test the impact of stellar parameters, and general observational uncertainties, among others. 

Of course, the use of stellar atmosphere models represents an obvious limitation to our results, rendering them subject to assumptions and simplifications inherent in the theoretical models. Any systematic errors, e.g. in temperature scale or in opacity tables, may influence the results presented here, and different atmospheric models may yield somewhat different results. In this context, the present work primarily seeks to provide guidance and a first estimation of the importance of \kcorr s for stellar standard candles and \Ho, notably for the TRGB stars and other potential evolved SSCs. However, any systematic offsets due to model assumptions should not invalidate the trends predicted by our analysis, e.g. as a function of temperature, redshift, reddening, metallicity, and photometric filters. We therefore believe that the key results of this work are insensitive to model inadequacies, notably regarding the Wesenheit function's natural mitigation of the magnitude of \kcorr s, the higher sensitivity of the F814W-based TRGB method to \kcorr s and missing corrections of host extinction, and the greater consistency of \kcorr s at longer wavelengths.

\subsection{Mira stars and the $J-$region AGB stars\label{disc:Mira}}

Short-period ($P \lesssim 400$\,d) oxygen-rich Mira stars have been used to measure distances to nearby galaxies \citep[e.g.][]{Whitelock2008,Yuan2017,Huang2018,Huang2020} and provide a potentially useful alternative to Cepheids or TRGB stars for calibrating SNeIa luminosity. Mira stars are typically M giants \citep{Yao2017} with spectral types varying between M5-M9 \citep[spectral classifications by \citealt{Sloan1998}]{Smith2002}, and the Mira prototype, $o$~Ceti, varies between M5IIIe - M9III \citep{Skiff2014}. The fraction of O- to C-rich Miras depends on metallicity, and most MW Mira stars re O-rich \citep[e.g.][]{Yao2017}. 

As dust-rich and cool AGB stars, Miras are clearly outside the parameter space covered by the ATLAS9 models and thus outside the scope of a detailed discussion. Nevertheless, the increasing importance of \kcorr s with decreasing \teff\ and increasing host extinction implies that the extinction-dependent \kcorr\ will be even more relevant for Mira stars than for Cepheids or TRGB stars. It would be interesting to check if this could be used to empirically constrain $K$ using differences in distance between Mira and Cepheid stars. Localized dust due to stellar mass-loss (CSEs) is likely to complicate very accurate host reddening corrections in part due to the poorly known intrinsic color of Mira stars at a given period. The quadratic form of the Mira LL is particularly sensitive to RLB \citep{Anderson2019rlb}. As is the case of Cepheids, the negative LL slope of Miras means that missing RLB corrections lead to systematic underestimates of \Ho.

\label{data:JAGB}Similarly, we could not directly investigate relativistic corrections for the recently introduced JAGB method \citep{Madore2020,Zgirski2021} since fairly little is currently known about these stars. However, we expect their cool temperatures and likely dusty envelopes to require \kcorr s more similar to Mira stars than to TRGB stars. Observing these stars with {\it JWST/NIRCAM} will have the benefit of smaller \kcorr s, although Wien's law increasingly shifts the SED maximum towards the F200W and F277W filters, which may increase sensitivity to extinction and $K$ in analogy with our results for the F814W TRGB method.

\subsection{Dust extinction\label{disc:dust}}

Dust extinction is the perennial bane of astronomical distance measurements. Aiming towards a $1\%$ \Ho\ measurement, we have explicitly incorporated the effect of dust extinction in non-comoving restframes in the computation of \kcorr s. The computed \kcorr s thus directly depend on the amount of host reddening assumed, and careful consideration of the expected host reddening is required when using the tabulated values of \Kext\ in Appendix\,\ref{app:tables}.

The Wesenheit function (cf. Sec.\,\ref{sec:reddening}) offers an easy-to-employ and robust opportunity for avoiding most of the  difficulties related to host reddening while simultaneously reducing the LL's intrinsic scatter \citep[e.g.][]{Macri2015,Anderson2016rot}. Additionally, we found here that \kcorr s computed using the optical-NIR Wesenheit function exhibit only a marginal redshift dependence thanks to the function's advantageous combination of passbands exhibiting opposite trends with redshift. Given these substantial benefits of the Wesenheit function, it appears judicious to also apply the Wesenheit function to TRGB distance measurements, although obtaining sufficiently precise observational material in multiple passbands represents a practical difficulty. A combination of optical and NIR filters would be particularly well-suited to this effect to sample both sides of TRGB star peak flux. 

As discussed by \citet{Moertsell2021}, the assumption of a particular reddening law normalized to 
a fixed value of $R_V$ remains a clear limitation of the Wesenheit formalism. Line-of-sight variations of {$R_V$ contribute uncertainty} when considering individual objects. However, the distance ladder probes hundreds of lines of sight among its calibrators located in multiple different anchor galaxies, as well as thousands of lines of sight among standard candles in SN-host galaxies \citep{Riess2021,Freedman2021}. 
Hence, we expect line-of-sight variations of $R_V$ to primarily contribute scatter to observed Wesenheit PL-distributions in anchor and host galaxies alike, rather than a systematic difference between anchor galaxies and SN-hosts. 

That said, cosmic expansion \emph{does} affect the apparent reddening of sources systematically as a function of redshift \citep{McCall2004}. However, this effect is on the order of a few milli-magnitudes for the near-IR Wesenheit function applied to distances shorter than $100$\,Mpc (cf. Eq.\,\ref{eq:Rofz}). The corresponding bias correction tends to increase the value of \Ho.

Another intriguing complication is the interplay of dust extinction with heterochromatic magnitudes measured in broadband filters. As discussed in Sec.\,\ref{sec:Wesenheit}, the value of $R^W$ in the Wesenheit function depends on the filter combination used, the adopted reddening law and its normalization $R_V$, and source SED. This latter effect of source SED on $R^W$ is relevant for the determination of \Ho\ \citep[cf. the analysis variants in][]{Riess2016}. However, it is of only secondary importance to our computed \kcorr s as long as $R_V$ and source SED are identical in the respective rest frames.

In summary, the Wesenheit function offers a straightforward and robust way to avoid significant issues related to dust extinction without explicitly correcting for reddening by measuring color excesses relative to somewhat uncertain fiducial colors. The selection of low-reddening Cepheids for LL calibration \citep{Riess2018,Riess2021} and low-reddening, nearly face-on SN-host galaxies \citet{Riess2016} for their application, as well as using color and amplitude cuts for Cepheids in SN-hosts further limits the impact of any possible negative side effects related to reddening. 
Currently missing host extinction corrections introduce a small (and uncertain) bias for  F814W-based TRGB distances. As explained in Sec\,\ref{data:TRGB}, two sources of extinction could contribute. Extinction by gas or dust in galaxy halos is mostly avoided by observing the TRGB far from the disk \citep{Jang2021}. Extinction due to CSEs has not yet been adequately quantified so that its impact on TRGB distances remains an open question. In particular \jwst's longer wavelength filters will reduce sensitivity to dust absorption, but may also increase sensitivity to warm circumstellar material. This, of course, also applies to Cepheids, where circumstellar environments have been reported in several studies \citep[e.g.][]{Kervella2009,Gallenne2013,Schmidt2015} and increasingly contribute flux at wavelengths longer than $\sim H-$band \citep[e.g.][]{Hocde2020,Groenewegen2020}.

\section{Conclusions\label{conclusions}}

We have investigated relativistic corrections for cosmic distances measured using stellar standard candles, such as classical Cepheids and stars near the TRGB. Our analysis focused on redshifts $z < 0.03$ to aid the accurate calibration of the SNeIa luminosity zero-point as required for measuring \Ho\ to within $1\%$. Three relativistic effects have been considered: \kcorr s, time-dilation, and the difference of dust extinction across non-comoving reference frames. To this end, we incorporated extinction explicitly in the definition of the \kcorr.

\kcorr s for the near-IR Wesenheit function applied to Cepheids slightly increase the value of \Ho\ previously reported by the SH0ES team \citep{Riess2016,Riess2018,Riess2021} and are relatively insensitive to redshift. 
Redshift-Leavitt bias corrections \citep{Anderson2019rlb} dominate over \Kw\ and are easily applied using the slope of the period-luminosity relation and redshift measurements of SN-host galaxies. The effect of redshift on the $R^W$ term for Wesenheit function is very small ($< 3$\,mmag). All three effects tend to increase the value of \Ho.

The combined $K-$, RLB, and Wesenheit slope correction increases the SH0ES value of \Ho\ by $0.45\pm0.05$\,\Hunit\ to $H_0^{\mathrm{SH0ES}} = 73.65 \pm 1.3$\,\Hunit, enhancing the tension with the value for the early Universe reported by the \citet{Planck2020H0} from $4.2 \sigma$ to $4.4 \sigma$. Larger corrections will apply as the average distance of SN-host galaxies is pushed farther in an effort to obtain larger samples of SN-host galaxies for better \Ho\ precision. 

\kcorr s for TRGB-based distances depend on the amount of localized host extinction. For little to no host reddening ($\mathrm{E(B-V)^{host}}\lesssim 0.004$\,mag), $K(\mathrm{F814W}) < 0$ for the TRGB fiducial SED considered. Reddening due to the circumgalactic medium equivalent to $\mathrm{E(B-V)^{host}}\approx 0.01$\,mag yields an upward correction of the TRGB-based \Ho\ measurement reported by \citet{Freedman2021} by $\sim 0.5\%$ to $H_0^{\mathrm{CCHP}} = 70.2 \pm 1.7$\,\Hunit. 
We identified reddening due to circumstellar material lost during RGB evolution as a potential additional contributor to host extinction, although it is unlikely to explain the difference between Cepheid and TRGB-based \Ho\ measurements. The impact of mass-loss on TRGB distance measurements remains an open question, and further research is required to understand its magnitude and relevance, in particular for longer-wavelength \jwst\ observations.

On the other hand, we estimate a combined correction of $-0.1$\,\Hunit\ applicable to the recent \Ho\ measurement by \citet{Anand2021H0} when taking into account both the \kcorr\ and the effect of redshift on the observed color used to correct population variations in the TRGB absolute magnitude. Thus, the TRGB methodology employed by the EDD \citep{Anand2021edd} yields a corrected value of $H_0^{\mathrm{EDD}} = 71.4\pm1.8$\,\Hunit, consistent to within $1\sigma$ with both $H_0^{\mathrm{SH0ES}}$ and $H_0^{\mathrm{CCHP}}$.

\kcorr s are required to measure \Ho\ to $1\%$ using all types of standard candles and using most photometric filters. Interestingly, \kcorr s become increasingly independent of \teff\ at wavelengths up to approximately $2\mu$m and exhibit the weakest \teff\ dependence in the \jwst\ F200W filter. Nevertheless, F200W \kcorr s for SSCs observed at Coma cluster distances (100\,Mpc) exceed $0.02$\,mag ($1\%$ in distance) for all temperatures considered. For TRGB stars, \jwst's F277W filter yields consistently small \kcorr s that do not exceed $1\%$ even at $100$\,Mpc. New Wesenheit functions formulated for a combination of shorter-wavelength \hst\ and longer-wavelength \jwst\ passbands could be designed to optimally reduce the combined effects of reddening and \kcorr s on distance measurements.

\begin{acknowledgements}
      I am grateful to Antoine M\'erand, Rolf-Peter Kudritzki, Laurent Eyer, Jason Spyromilio, and Bruno Leibundgut for inspiring and useful discussions. Comments by the anonymous referee resulted in an improved manuscript, in particular regarding the discussion of host reddening and TRGB distances. I also thank the ORIGINS excellence cluster and the Universit\"atssternwarte of the Ludwig-Maximilians-Universit\"at M\"unchen, Germany, for hosting me as a long-term visitor during uncertain times related to the COVID-19 pandemic in 2020. 
      
      This research has made use of the SVO Filter Profile Service (\url{http://svo2.cab.inta-csic.es/theory/fps/}) supported from the Spanish MINECO through grant AYA2017-84089 and the NASA/IPAC Extragalactic Database (NED), which is operated by the Jet Propulsion Laboratory, California Institute of Technology, under contract with the National Aeronautics and Space Administration.

      The following software made this research possible: \texttt{Pysynphot}, \texttt{astropy} \citep{astropy2013,astropy2018}, \texttt{ipython} \citep{ipython}, \texttt{numpy} \citep{numpy}, \texttt{matplotlib} \citep{matplotlib}, and \texttt{scipy} \citep{scipy}. 

      This project has received funding from the European Research Council (ERC) under the European Union's Horizon 2020 research and innovation programme (Grant Agreement No. 947660).

      RIA acknowledges funding provided by an SNSF Eccellenza Professorial Fellowship (Grant No. PCEFP2\_194638).
\end{acknowledgements}

\bibliographystyle{aa} 
\bibliography{biblio}

\begin{appendix}
\section{Information provided in machine-readable tables\label{app:tables}}

This appendix presents a subset of the information made available as machine-readable tables via the Centre de Donn\'ees de Strasbourg for reference. Table\,\ref{app:tabgrid} illustrates the information available for the model grid computation, assuming that absolute calibration accounts for extinction and that foreground reddening corrections are also applied, i.e., $\mathrm{E(B-V)^{cal}} = \mathrm{E(B-V)^{fg}} = 0$. Table\,\ref{app:WesenCepheid} lists the three corrective terms, \Kw, RLB, and $\Delta \mu^{R^W}$ for the fiducial Cepheid observed using the optical-NIR Wesenheit function (Eq.\,\ref{eq:Wesenheit}). Table\,\ref{app:TRGBebv} lists results for stars near the TRGB for a finely-sampled, narrow range of host reddening values.

\begin{sidewaystable*}
\caption{\label{app:tabgrid}$K-$corrections as a function of stellar parameters, redshift, and reddening.}
\begin{tabular}{@{}c@{\hspace{4pt}}c@{\hspace{4pt}}c@{\hspace{4pt}}c@{\hspace{4pt}}c@{\hspace{4pt}}|@{\hspace{4pt}}r@{\hspace{6pt}}r@{\hspace{4pt}}|@{\hspace{4pt}}r@{\hspace{6pt}}r@{\hspace{4pt}}|@{\hspace{4pt}}r@{\hspace{6pt}}r@{\hspace{4pt}}|@{\hspace{4pt}}r@{\hspace{6pt}}r@{\hspace{4pt}}|@{\hspace{4pt}}r@{\hspace{6pt}}r@{\hspace{4pt}}|@{\hspace{4pt}}r@{\hspace{6pt}}r@{\hspace{4pt}}|@{\hspace{4pt}}r@{\hspace{6pt}}r@{\hspace{4pt}}|@{\hspace{4pt}}r@{\hspace{6pt}}r@{\hspace{4pt}}|@{\hspace{4pt}}r@{\hspace{6pt}}r@{}}
\hline
\teff\ & \logg & [Fe/H] & $z$ & E(B-V) & \multicolumn{2}{c}{F555W} & \multicolumn{2}{c}{F814W} & \multicolumn{2}{c}{F160W} & \multicolumn{2}{c}{2MASS $J$} & \multicolumn{2}{c}{2MASS $K_s$} & \multicolumn{2}{c}{F115W} & \multicolumn{2}{c}{F150W} & \multicolumn{2}{c}{F200W} & \multicolumn{2}{c}{F277W}  \\
 & & & & & \Ktab & \Aapp & \Ktab & \Aapp & \Ktab & \Aapp & \Ktab & \Aapp & \Ktab & \Aapp & \Ktab & \Aapp & \Ktab & \Aapp & \Ktab & \Aapp & \Ktab & \Aapp \\ 
 (K) & (cgs) & & & (mag) & (mag)& (mag)& (mag)& (mag)& (mag)& (mag)& (mag)& (mag)& (mag)& (mag)& (mag)& (mag)& (mag)& (mag)& (mag)& (mag) & (mag) & (mag) \\
\hline
5400 & 1.5 & 0.00 & 0.005 & 0.00 & -0.010 & 0.000 & -0.001 & 0.000 & -0.001 & 0.000 & -0.002 & 0.000 &  0.005 & 0.000 & -0.002 & 0.000 & -0.001 & 0.000 &  0.005 & 0.000 &  0.004 & 0.000 \\
5400 & 1.5 & 0.00 & 0.005 & 0.01 &  0.022 & 0.032 &  0.017 & 0.018 &  0.005 & 0.006 &  0.007 & 0.009 &  0.009 & 0.004 &  0.008 & 0.010 &  0.006 & 0.006 &  0.009 & 0.004 &  0.008 & 0.004 \\
5400 & 1.5 & 0.00 & 0.005 & 0.03 &  0.086 & 0.095 &  0.054 & 0.054 &  0.018 & 0.018 &  0.024 & 0.026 &  0.016 & 0.011 &  0.028 & 0.029 &  0.019 & 0.019 &  0.018 & 0.013 &  0.015 & 0.011 \\
5400 & 1.5 & 0.00 & 0.005 & 0.05 &  0.151 & 0.158 &  0.091 & 0.089 &  0.031 & 0.031 &  0.042 & 0.043 &  0.024 & 0.018 &  0.048 & 0.049 &  0.032 & 0.032 &  0.026 & 0.021 &  0.022 & 0.018 \\
5400 & 1.5 & 0.00 & 0.005 & 0.10 &  0.311 & 0.315 &  0.183 & 0.179 &  0.062 & 0.061 &  0.087 & 0.085 &  0.042 & 0.035 &  0.099 & 0.097 &  0.066 & 0.064 &  0.048 & 0.042 &  0.039 & 0.035 \\
5400 & 1.5 & 0.00 & 0.010 & 0.00 & -0.021 & 0.000 & -0.003 & 0.000 & -0.002 & 0.000 & -0.003 & 0.000 &  0.010 & 0.000 & -0.003 & 0.000 & -0.001 & 0.000 &  0.010 & 0.000 &  0.009 & 0.000 \\
5400 & 1.5 & 0.00 & 0.010 & 0.01 &  0.011 & 0.032 &  0.015 & 0.018 &  0.004 & 0.006 &  0.006 & 0.009 &  0.014 & 0.004 &  0.007 & 0.010 &  0.005 & 0.006 &  0.014 & 0.004 &  0.012 & 0.004 \\
5400 & 1.5 & 0.00 & 0.010 & 0.03 &  0.075 & 0.095 &  0.052 & 0.054 &  0.017 & 0.018 &  0.023 & 0.026 &  0.021 & 0.011 &  0.027 & 0.029 &  0.018 & 0.019 &  0.023 & 0.013 &  0.019 & 0.011 \\
5400 & 1.5 & 0.00 & 0.010 & 0.05 &  0.139 & 0.157 &  0.089 & 0.089 &  0.030 & 0.031 &  0.041 & 0.043 &  0.029 & 0.018 &  0.047 & 0.049 &  0.032 & 0.032 &  0.031 & 0.021 &  0.026 & 0.018 \\
5400 & 1.5 & 0.00 & 0.010 & 0.10 &  0.299 & 0.315 &  0.180 & 0.178 &  0.061 & 0.061 &  0.085 & 0.085 &  0.047 & 0.035 &  0.097 & 0.097 &  0.064 & 0.064 &  0.053 & 0.042 &  0.044 & 0.035 \\
5400 & 1.5 & 0.00 & 0.015 & 0.00 & -0.031 & 0.000 & -0.004 & 0.000 & -0.002 & 0.000 & -0.004 & 0.000 &  0.016 & 0.000 & -0.005 & 0.000 & -0.003 & 0.000 &  0.015 & 0.000 &  0.013 & 0.000 \\
5400 & 1.5 & 0.00 & 0.015 & 0.01 &  0.001 & 0.031 &  0.014 & 0.018 &  0.004 & 0.006 &  0.005 & 0.009 &  0.019 & 0.004 &  0.005 & 0.010 &  0.004 & 0.006 &  0.020 & 0.004 &  0.017 & 0.004 \\
5400 & 1.5 & 0.00 & 0.015 & 0.03 &  0.065 & 0.094 &  0.050 & 0.053 &  0.016 & 0.018 &  0.023 & 0.026 &  0.027 & 0.010 &  0.025 & 0.029 &  0.017 & 0.019 &  0.028 & 0.013 &  0.024 & 0.011 \\
5400 & 1.5 & 0.00 & 0.015 & 0.05 &  0.128 & 0.157 &  0.087 & 0.089 &  0.029 & 0.030 &  0.040 & 0.043 &  0.034 & 0.017 &  0.045 & 0.049 &  0.030 & 0.032 &  0.036 & 0.021 &  0.031 & 0.018 \\
5400 & 1.5 & 0.00 & 0.015 & 0.10 &  0.287 & 0.314 &  0.177 & 0.178 &  0.060 & 0.061 &  0.084 & 0.085 &  0.052 & 0.035 &  0.094 & 0.097 &  0.063 & 0.064 &  0.058 & 0.042 &  0.048 & 0.035 \\
5400 & 1.5 & 0.00 & 0.020 & 0.00 & -0.040 & 0.000 & -0.005 & 0.000 & -0.003 & 0.000 & -0.007 & 0.000 &  0.021 & 0.000 & -0.007 & 0.000 & -0.004 & 0.000 &  0.020 & 0.000 &  0.017 & 0.000 \\
5400 & 1.5 & 0.00 & 0.020 & 0.01 & -0.009 & 0.031 &  0.013 & 0.018 &  0.003 & 0.006 &  0.002 & 0.009 &  0.024 & 0.003 &  0.003 & 0.010 &  0.002 & 0.006 &  0.024 & 0.004 &  0.021 & 0.004 \\
5400 & 1.5 & 0.00 & 0.020 & 0.03 &  0.055 & 0.094 &  0.049 & 0.053 &  0.016 & 0.018 &  0.019 & 0.026 &  0.032 & 0.010 &  0.022 & 0.029 &  0.015 & 0.019 &  0.033 & 0.012 &  0.028 & 0.011 \\
5400 & 1.5 & 0.00 & 0.020 & 0.05 &  0.118 & 0.157 &  0.085 & 0.089 &  0.028 & 0.030 &  0.037 & 0.042 &  0.039 & 0.017 &  0.042 & 0.048 &  0.028 & 0.032 &  0.041 & 0.021 &  0.035 & 0.018 \\
5400 & 1.5 & 0.00 & 0.020 & 0.10 &  0.275 & 0.313 &  0.175 & 0.177 &  0.059 & 0.061 &  0.080 & 0.085 &  0.057 & 0.035 &  0.091 & 0.097 &  0.061 & 0.064 &  0.062 & 0.042 &  0.052 & 0.035 \\
5400 & 1.5 & 0.00 & 0.025 & 0.00 & -0.049 & 0.000 & -0.006 & 0.000 & -0.004 & 0.000 & -0.009 & 0.000 &  0.025 & 0.000 & -0.010 & 0.000 & -0.005 & 0.000 &  0.024 & 0.000 &  0.021 & 0.000 \\
5400 & 1.5 & 0.00 & 0.025 & 0.01 & -0.018 & 0.031 &  0.012 & 0.018 &  0.003 & 0.006 & -0.001 & 0.008 &  0.029 & 0.003 &  0.000 & 0.010 &  0.001 & 0.006 &  0.028 & 0.004 &  0.025 & 0.004 \\
5400 & 1.5 & 0.00 & 0.025 & 0.03 &  0.045 & 0.094 &  0.048 & 0.053 &  0.015 & 0.018 &  0.016 & 0.025 &  0.036 & 0.010 &  0.020 & 0.029 &  0.014 & 0.019 &  0.037 & 0.012 &  0.032 & 0.010 \\
5400 & 1.5 & 0.00 & 0.025 & 0.05 &  0.108 & 0.156 &  0.083 & 0.089 &  0.027 & 0.030 &  0.034 & 0.042 &  0.043 & 0.017 &  0.039 & 0.048 &  0.027 & 0.032 &  0.045 & 0.021 &  0.039 & 0.017 \\
5400 & 1.5 & 0.00 & 0.025 & 0.10 &  0.264 & 0.312 &  0.172 & 0.177 &  0.058 & 0.061 &  0.077 & 0.085 &  0.062 & 0.035 &  0.088 & 0.097 &  0.059 & 0.064 &  0.066 & 0.041 &  0.056 & 0.035 \\
5400 & 1.5 & 0.00 & 0.030 & 0.00 & -0.059 & 0.000 & -0.007 & 0.000 & -0.006 & 0.000 & -0.012 & 0.000 &  0.030 & 0.000 & -0.012 & 0.000 & -0.006 & 0.000 &  0.028 & 0.000 &  0.025 & 0.000 \\
5400 & 1.5 & 0.00 & 0.030 & 0.01 & -0.027 & 0.031 &  0.011 & 0.018 &  0.001 & 0.006 & -0.003 & 0.008 &  0.034 & 0.003 & -0.002 & 0.010 &  0.000 & 0.006 &  0.032 & 0.004 &  0.029 & 0.003 \\
5400 & 1.5 & 0.00 & 0.030 & 0.03 &  0.035 & 0.094 &  0.046 & 0.053 &  0.013 & 0.018 &  0.014 & 0.025 &  0.041 & 0.010 &  0.017 & 0.029 &  0.013 & 0.019 &  0.040 & 0.012 &  0.036 & 0.010 \\
5400 & 1.5 & 0.00 & 0.030 & 0.05 &  0.097 & 0.156 &  0.081 & 0.088 &  0.025 & 0.030 &  0.031 & 0.042 &  0.048 & 0.017 &  0.036 & 0.048 &  0.026 & 0.032 &  0.048 & 0.021 &  0.043 & 0.017 \\
5400 & 1.5 & 0.00 & 0.030 & 0.10 &  0.253 & 0.312 &  0.170 & 0.177 &  0.055 & 0.060 &  0.074 & 0.084 &  0.066 & 0.035 &  0.085 & 0.096 &  0.058 & 0.064 &  0.069 & 0.041 &  0.060 & 0.035 \\
\multicolumn{23}{c}{\ldots}\\
\hline
\end{tabular}
\tablefoot{Extinction-dependent \kcorr s, \Kext, for a wide parameter range covering \teff\ between $3500$ and $6000$\,K; \logg\ between $2.0$ and $0.0$; [Fe/H] between $-2.0$ and $0.5$; redshift $z$ between $0.005$ and $0.030]$, and host reddening $\mathrm{E(B-V)^{host}}$ between $0.0$ and $0.5$\,mag in the {\it HST/WFC3} filters F555W, F814W, and F160W, 2MASS J \& $K_s$, and {\it JWST/NIRCAM} filters F115W, F150W, F200W, and F277W.
\Kext\ is listed together with $A^{\rm{app}}$ (extinction correction based on apparent color excess) for each filter; \Kred$=$\Kext$-A^{\rm{app}}$ (Eq.\,\ref{eq:Kred}). Extinction applicable to absolute calibration and the foreground is assumed to be corrected, that is, $\mathrm{E(B-V)^{cal}} = \mathrm{E(B-V)^{fg}}=0.00$\,mag. The subset of results shown highlight results for the fiducial 10\,d Cepheid. The full version of the table is available in machine-readable format at the CDS.}
\end{sidewaystable*}

\begin{table}
\caption{\label{app:WesenCepheid}Relativistic corrections for fiducial 10\,d Cepheids observed using Wesenheit magnitudes}
\centering
\begin{tabular}{@{}c@{\hspace{6pt}}c@{\hspace{6pt}}c@{\hspace{6pt}}c@{\hspace{6pt}}c@{\hspace{6pt}}r@{\hspace{6pt}}r@{\hspace{6pt}}r@{}}
\hline
\teff\ & \logg & [Fe/H] & $z$ & E(B-V) & \Kw & $\Delta\mu^{\mathrm{RLB}}$ & $\Delta\mu^{R^W}$ \\
(K) & (cgs) & & & (mag) & (mmag) & (mmag) & (mmag) \\
\hline
5400 & 1.5 & 0.00 & 0.005 & 0.00 &  2 &  7 &  1 \\
5400 & 1.5 & 0.00 & 0.005 & 0.01 &  2 &  7 &  1 \\
5400 & 1.5 & 0.00 & 0.005 & 0.03 &  2 &  7 &  1 \\
5400 & 1.5 & 0.00 & 0.005 & 0.05 &  2 &  7 &  1 \\
5400 & 1.5 & 0.00 & 0.005 & 0.10 &  3 &  7 &  1 \\
5400 & 1.5 & 0.00 & 0.005 & 0.20 &  4 &  7 &  1 \\
5400 & 1.5 & 0.00 & 0.005 & 0.30 &  6 &  7 &  1 \\
5400 & 1.5 & 0.00 & 0.005 & 0.40 &  8 &  7 &  1 \\
5400 & 1.5 & 0.00 & 0.005 & 0.50 & 10 &  7 &  1 \\
5400 & 1.5 & 0.00 & 0.025 & 0.00 &  8 & 35 &  3 \\
5400 & 1.5 & 0.00 & 0.025 & 0.01 &  8 & 35 &  3 \\
5400 & 1.5 & 0.00 & 0.025 & 0.03 &  8 & 35 &  3 \\
5400 & 1.5 & 0.00 & 0.025 & 0.05 &  8 & 35 &  3 \\
5400 & 1.5 & 0.00 & 0.025 & 0.10 &  8 & 35 &  3 \\
5400 & 1.5 & 0.00 & 0.025 & 0.20 &  8 & 35 &  3 \\
5400 & 1.5 & 0.00 & 0.025 & 0.30 &  8 & 35 &  3 \\
5400 & 1.5 & 0.00 & 0.025 & 0.40 &  8 & 35 &  3 \\
5400 & 1.5 & 0.00 & 0.025 & 0.50 &  9 & 35 &  3 \\
\multicolumn{6}{c}{\ldots} & \\
\hline
\end{tabular}
\tablefoot{Extinction-dependent Wesenheit function \kcorr s, \Kw, redshift-Leavitt bias corrections (Eq.\,\ref{eq:RLB}), $\Delta \mu^{\mathrm{RLB}}$, and the Wesenheit slope correction (Eq.\,\ref{eq:deltaMuRw}), $\Delta \mu^{R^W}$, for the fiducial 10\,d Cepheid assuming $\mathrm{E(B-V)^{cal}} = 0.4$\,mag and $\mathrm{E(B-V)^{fg}} = 0.0223$\,mag, cf. Sec.\,\ref{data:Cepheids}. The full version of this table is available via the CDS.}
\end{table}

\begin{table*}
\caption{\label{app:TRGBebv}$K-$corrections for TRGB stars}
\centering
\begin{tabular}{@{}c@{\hspace{4pt}}c@{\hspace{4pt}}c@{\hspace{4pt}}c@{\hspace{4pt}}c@{\hspace{4pt}}c@{\hspace{4pt}}c@{\hspace{4pt}}c@{\hspace{4pt}}c@{\hspace{4pt}}c@{\hspace{4pt}}c@{\hspace{4pt}}c@{\hspace{4pt}}c@{\hspace{4pt}}c@{\hspace{4pt}}c@{\hspace{4pt}}c@{\hspace{4pt}}c@{}}
\hline
\rule{0pt}{2.6ex} & & & & & \multicolumn{4}{c}{{\it HST/WFC3}} & \multicolumn{2}{c}{\it HST/ACS} & \multicolumn{2}{c}{2MASS} & \multicolumn{4}{c}{\it JWST/NIRCAM} \\
\teff & \logg & [Fe/H] & $z$ & E(B-V) & $W_{\mathrm{H,VI}}$ & F555W & F814W & F160W & F606W & F814W & J & K$_s$ & F115W & F150W & F200W & F277W \\
(K) & (cgs) & & & (mag) & \multicolumn{12}{c}{(mmag)}  \rule[-0.9ex]{0pt}{0pt}\\
\hline
4200 & 0.5 & -1.75 & 0.005 & 0.000 &  -2 & -21 &  -7 &  -8 & -16 &  -8 &  -7 &   6 &  -6 &  -7 &   5 &   \rule{0pt}{2.6ex}3  \\
4200 & 0.5 & -1.75 & 0.005 & 0.002 &  -2 & -15 &  -4 &  -7 & -11 &  -4 &  -5 &   7 &  -4 &  -6 &   5 &   4 \\
4200 & 0.5 & -1.75 & 0.005 & 0.004 &  -2 &  -9 &   0 &  -6 &  -5 &  -0 &  -4 &   7 &  -2 &  -5 &   6 &   4 \\
4200 & 0.5 & -1.75 & 0.005 & 0.006 &  -2 &  -2 &   4 &  -4 &   0 &   3 &  -2 &   8 &  -0 &  -4 &   7 &   5 \\
4200 & 0.5 & -1.75 & 0.005 & 0.008 &  -2 &   4 &   7 &  -3 &   6 &   7 &  -0 &   9 &   2 &  -2 &   8 &   6 \\
4200 & 0.5 & -1.75 & 0.005 & 0.010 &  -2 &  10 &  11 &  -2 &  12 &  11 &   2 &  10 &   4 &  -1 &   9 &   6 \\
4200 & 0.5 & -1.75 & 0.005 & 0.015 &  -1 &  26 &  20 &   1 &  25 &  20 &   6 &  11 &   9 &   2 &  11 &   8 \\
4200 & 0.5 & -1.75 & 0.005 & 0.020 &  -1 &  42 &  29 &   4 &  39 &  29 &  10 &  13 &  13 &   6 &  13 &  10 \\
4200 & 0.5 & -1.75 & 0.005 & 0.025 &  -1 &  57 &  38 &   7 &  53 &  38 &  15 &  15 &  18 &   9 &  15 &  12 \\
4200 & 0.5 & -1.75 & 0.005 & 0.030 &  -1 &  73 &  47 &  11 &  67 &  47 &  19 &  17 &  23 &  12 &  17 &  13 \\
4200 & 0.5 & -1.75 & 0.005 & 0.040 &  -1 & 104 &  66 &  17 &  94 &  65 &  28 &  21 &  33 &  19 &  22 &  17 \\
4200 & 0.5 & -1.75 & 0.005 & 0.050 &  -1 & 135 &  84 &  23 & 122 &  84 &  37 &  24 &  43 &  25 &  26 &  20 \\
4200 & 0.5 & -1.75 & 0.010 & 0.000 &  -4 & -42 & -15 & -16 & -32 & -15 & -13 &  11 & -12 & -15 &  10 &   6 \\
4200 & 0.5 & -1.75 & 0.010 & 0.002 &  -4 & -36 & -11 & -15 & -27 & -11 & -11 &  12 & -10 & -14 &  10 &   7 \\
4200 & 0.5 & -1.75 & 0.010 & 0.004 &  -3 & -30 &  -7 & -14 & -21 &  -7 &  -9 &  13 &  -8 & -13 &  11 &   7 \\
4200 & 0.5 & -1.75 & 0.010 & 0.006 &  -3 & -24 &  -4 & -13 & -16 &  -4 &  -7 &  13 &  -6 & -11 &  12 &   8 \\
4200 & 0.5 & -1.75 & 0.010 & 0.008 &  -3 & -18 &  -0 & -11 & -10 &  -0 &  -6 &  14 &  -4 & -10 &  13 &   9 \\
4200 & 0.5 & -1.75 & 0.010 & 0.010 &  -3 & -11 &   3 & -10 &  -5 &   3 &  -4 &  15 &  -2 &  -9 &  14 &   9 \\
4200 & 0.5 & -1.75 & 0.010 & 0.015 &  -3 &   4 &  12 &  -7 &   9 &  12 &   1 &  17 &   3 &  -5 &  16 &  11 \\
4200 & 0.5 & -1.75 & 0.010 & 0.020 &  -3 &  20 &  22 &  -4 &  23 &  22 &   5 &  18 &   8 &  -2 &  18 &  13 \\
4200 & 0.5 & -1.75 & 0.010 & 0.025 &  -3 &  35 &  31 &  -1 &  36 &  31 &   9 &  20 &  13 &   1 &  20 &  15 \\
4200 & 0.5 & -1.75 & 0.010 & 0.030 &  -3 &  51 &  40 &   2 &  50 &  40 &  14 &  22 &  18 &   4 &  22 &  16 \\
4200 & 0.5 & -1.75 & 0.010 & 0.040 &  -3 &  82 &  58 &   9 &  77 &  58 &  23 &  26 &  27 &  11 &  27 &  20 \\
4200 & 0.5 & -1.75 & 0.010 & 0.050 &  -3 & 113 &  76 &  15 & 105 &  76 &  31 &  29 &  37 &  17 &  31 &  \rule[-0.9ex]{0pt}{0pt}23  \\
\multicolumn{17}{c}{\ldots} \\
\hline
\end{tabular}
\tablefoot{Extinction-dependent \kcorr s, \Kext, applicable to stars near the TRGB for the restricted ranges of $T_{\mathrm{eff}}$ between 3800 and 4500\,K, $\log{g} \in [0.0,0.5]$, $\mathrm{[Fe/H]} \in [-2.0,-1.75,-1.5]$, and E(B-V) between $0.000$ and $0.050$. Extinction applicable to absolute calibration and the foreground is assumed to be corrected, that is, $\mathrm{E(B-V)^{cal}} = \mathrm{E(B-V)^{fg}}=0.00$\,mag. The full version of this table is available via the CDS.}
\end{table*}

\end{appendix}

\end{document}